\documentclass{vldb}
\pdfoutput=1

\RequirePackage{sjvanschaik-paper}

\input{settings.tex}

\usepackage{hyperref}
\hypersetup{
    pdfauthor={Sebastiaan J. van Schaik, Dan Olteanu, Robert Fink},
    pdftitle={ENFrame: A Platform for Processing Probabilistic Data},
    pdfkeywords={probabilistic data, data mining, clustering, frequent item sets, 
      hierarchical clustering, k-medoids, k-means, parallel computation, provenance,
      probability computation, SPROUT, MayBMS}
    pdfsubject={Probabilistic Data Mining}
}

\begin{document}
\clubpenalty=10000
\widowpenalty = 10000

\title{ENFrame: A Platform for Processing Probabilistic Data}

\numberofauthors{3}
\author{%
\alignauthor \mbox{Sebastiaan J. van Schaik$^{1,2}$}
\alignauthor Dan Olteanu$^{1}$
\alignauthor Robert Fink$^{1}$\and
% add some space between author names and affils
\fontsize{10}{10}\selectfont\ttfamily\upshape
\{Robert.Fink,Dan.Olteanu,Sebastiaan.van.Schaik\}@cs.ox.ac.uk\vspace*{0.5mm}\\
\fontsize{10}{10}\selectfont\itshape $~^{1}$Department of Computer Science, University of Oxford, United Kingdom\vspace*{0.5mm}\\
\fontsize{10}{10}\selectfont\itshape $~^{2}$Oxford e-Research Centre, University of Oxford, United Kingdom
}

\maketitle
\begin{abstract}
  This paper introduces ENFrame, a unified data processing platform
  for querying and mining probabilistic data. Using ENFrame, users can
  write programs in a fragment of Python with constructs such as
  bounded-range loops, list comprehension, aggregate operations on
  lists, and calls to external database engines. The program is then
  interpreted probabilistically by ENFrame.

  The realisation of ENFrame required novel contributions along
  several directions. We propose an event language that is expressive
  enough to succinctly encode arbitrary correlations, trace the
  computation of user programs, and allow for computation of discrete
  probability distributions of program variables. We exemplify ENFrame
  on three clustering algorithms: $k$-means, $k$-medoids, and Markov
  Clustering. We introduce sequential and distributed algorithms for
  computing the probability of interconnected events exactly or
  approximately with error guarantees.

  Experiments with \kmedoids clustering of sensor readings from energy
  networks show orders-of-magnitude improvements of exact clustering using ENFrame over na\"ive
  clustering in each possible world, of approximate over exact, and of
  distributed over sequential algorithms.
\end{abstract}

\nop{
}

\section{Introduction}
\label{sec:introduction}

Recent years have witnessed a solid body of work in probabilistic
databases with sustained systems building effort and extensive
analysis of computational problems for rich classes of queries and
probabilistic data models of varying expressivity~\cite{SORK:2011}. In
contrast, most state-of-the-art probabilistic data mining approaches
so far consider the restricted model of probabilistically independent
input and produce hard, deterministic
output~\cite{Aggarwal:2009}. This technology gap hinders the
development of data processing systems that integrate techniques for
both probabilistic databases and data mining.

The ENFrame data processing platform aims at closing this gap
by allowing users to specify iterative programs to query and mine
probabilistic data. The semantics of ENFrame programs is based on a unified
probabilistic interpretation of the entire processing pipeline from the input
data to the program result. It features an expressive set of programming constructs,
such as assignments, bounded-range loops, list comprehension, and aggregate
operations on lists, and calls to external database engines, coupled with
aspects of probabilistic databases, such as possible worlds semantics, arbitrary
data correlations, and exact and approximate probability computation with error
guarantees. The existing probabilistic data mining algorithms do not share these
latter aspects.

Under the \emph{possible worlds semantics}, the input is a probability
distribution over a finite set of possible worlds, whereby each world defines a
standard database or a set of input data points. The result of a user program is
equivalent to executing it within each world and is thus a probability
distribution over possible outcomes (e.g., partitionings). ENFrame exploits the fact
that many of the possible worlds are alike, and avoids iterating over the
exponentially many worlds.

Correlations occur naturally in query results~\cite{SORK:2011}, after
conditioning probabilistic databases using constraints~\cite{KO2008},
and are supported by virtually all mainstream probabilistic models. If correlations are ignored, the output can be arbitrarily off from the
expected result~\cite{Volk:2009:ClusteringWorlds,Aggarwal:2013:clusteringchapter}. For instance, consider two similar, but contradicting
sensor readings (mutually exclusive data points) in a clustering setting. There is no
possible world and thus no cluster containing both points, yet by ignoring
their negative correlation, we would assign them to the same cluster.

\begin{figure*}
  \begin{minipage}[t]{0.55\textwidth}
    \begin{scriptsize}
      \begin{verbatim}
1:  (O, n) = loadData()        # list and number of objects
2:  (k, iter) = loadParams()   # number of clusters and iterations
3:  M = init()                 # initialise medoids

4:  for it in range(0,iter):   # clustering iterations
5:   InCl = [None] * k         # assignment phase
6:   for i in range(0,k):
7:    InCl[i] = [None] * n
8:    for l in range(0,n):
9:     InCl[i][l] = reduce_and(
10:       [(dist(O[l],M[i]) <= dist(O[l],M[j])) for j in range(0,k)])
11:  InCl = breakTies2(InCl)   # each object is in exactly one cluster

12:  DistSum = [None] * k      # update phase
13:  for i in range(0,k):
14:   DistSum[i] = [None] * n
15:   for l in range(0,n):
16:    DistSum[i][l] = reduce_sum(
17:      [dist(O[l],O[p]) for p in range(0,n) if InCl[i][p]])

18:  Centre = [None] * k
19:  for i in range(0,k):
20:   Centre[i] = [None] * n
21:   for l in range(0,n):
22:    Centre[i][l] = reduce_and(
23:      [DistSum[i][l] <= DistSum[i][p] for p in range(0,n)])
24:  Centre = breakTies1(Centre)  # enforce one Centre per cluster

25:  M = [None] * k
26:  for i in range(0,k):
27:   M[i] = reduce_sum([O[l] for l in range(0,n) if Centre[i][l]])
      \end{verbatim}
    \end{scriptsize}
  \end{minipage}
  \begin{minipage}[t]{0.45\textwidth}
    \begin{scriptsize}
      \hspace*{0em}$\forall i \text{ in } 0..n-1: O^i \equiv \Phi(o_i) \otimes \vec{o}_i$\\
      \hspace*{0em}$M^0_{-1} \equiv \Phi(o_{\pi(0)}) \otimes \vec{o}_{\pi(0)}; \ldots; M^{k-1}_{-1} \equiv \Phi(o_{\pi(k-1)}) \otimes \vec{o}_{\pi(k-1)}$\\

      \vspace*{0.8em}
      \hspace*{0em}$\forall \text{it} \text{ in } 0..\text{iter}-1:$\\
      \hspace*{0.7em}$\forall i \text{ in } 0..k-1:$\\
      \hspace*{1.4em}$\forall l \text{ in } 0..n-1:$\\
      \hspace*{2.1em}$\text{InCl}^{i,l}_{\text{it}} \equiv \bigwedge_{j=0}^{k-1} \bigl[
          \dist(O^l, M^i_{\text{it}-1}) \leq \dist(O^l, M^j_{\text{it}-1})\bigr]$\\

      \vspace*{2.1em}
      \hspace*{0.7em}\# Encoding of \texttt{breakTies2} omitted

      \vspace*{0.9em}
      \hspace*{0.7em}$\forall i \text{ in } 0..k-1:$\\
      \hspace*{1.4em}$\forall l \text{ in } 0..n-1:$\\
      \hspace*{2.1em}$\text{DistSum}^{i,l}_{\text{it}} \equiv 
          \sum_{p=0}^{n-1} \text{InCl}^{i,p}_{\text{it}} \wedge \top\otimes\text{dist}(O^l, O^p)$\\

      \vspace*{3.1em}
      \hspace*{0.7em}$\forall i \text{ in } 0..k-1:$\\
      \hspace*{1.4em}$\forall l \text{ in } 0..n-1:$\\
      \hspace*{2.1em}$\text{Centre}^{i,l}_{\text{it}} \equiv 
          \bigwedge_{p=0}^{n-1} \bigl[ \text{DistSum}^{i,l}_{\text{it}} \leq \text{DistSum}^{i,p}_{\text{it}} \bigr]$\\

      \vspace*{2.1em}
      \hspace*{0.7em}\# Encoding of \texttt{breakTies2} omitted

      \vspace*{0.9em}
      \hspace*{0.7em}$\forall i \text{ in } 0..k-1:$\\
      \hspace*{1.4em}$M^i_{\text{it}} = \sum_{l=0}^{n-1} \text{Centre}^{i,l}_{\text{it}} \wedge O^l$
    \end{scriptsize}
  \end{minipage}%
  \vspace*{-4mm}%
  \caption{K-medoids clustering specified as user program (left) and simplified event program (right).\label{fig:KMedoidsUserProgram}}%
  \vspace*{-5mm}
\end{figure*}

The user is oblivious to the probabilistic nature of the input data,
and can write programs as if the input data were deterministic. It is
the task of ENFrame to interpret the program probabilistically.  The
approach taken here is to trace the user computation using
fine-grained provenance, which we call \emph{events}.  The event
language is a non-trivial extension of provenance
semirings~\cite{Green-2007} and semimodules~\cite{Amsterdamer2011}
that are used to trace the evaluation of positive relational algebra
queries with aggregates and to compute probabilities of query
results~\cite{FHO:VLDB12}. It features events with negation,
aggregates, loops, and definitions. It is expressive enough to
succinctly encode arbitrary correlations occurring in the input data
(e.g., modelled on Bayesian networks and pc-tables) and in the result
of the user program (e.g., co-occurrence of data points in the same
cluster), and trace the program state at any time. By annotating each
computation in the program with events, we effectively translate it into
an event program: variables become random variables whose possible
outcomes are conditioned on events. Selected events represent the probabilistic program output, \eg in case of clustering: the probability that a data point is a medoid, or the probability that two data points are assigned to the same cluster.  Besides probability computation,
events can be used for sensitivity analysis and explanation of the
program result.

The most expensive task supported by ENFrame is probability
computation for event programs, which is \#P-hard in general. We
developed sequential and distributed algorithms for both exact and
approximate probability computation with error guarantees. The
algorithms operate on a graph representations of the event programs
called \emph{event networks}. Expressions common to several events are
only represented once in such graphs. Event networks for data mining
tasks are very repetitive and highly interconnected due to the
combinatorial nature of the algorithms: the events at each iteration
are expressions over the events at the previous iteration and have the
same structure at each iteration. Moreover, the event networks can be
cyclic, so as to account for program loops. While it is possible to
unfold bounded-range loops, this can lead to prohibitively large event
networks.

The key challenge faced by ENFrame is to compute the probabilities of
a large number of interconnected events that are defined over several
iterations. Rather than computing the probability of each event separately, ENFrame's algorithms employ a novel \emph{bulk-compilation} technique, using Shannon expansion to depth-first explore the decision tree induced by the input random variables and the events in the program. The approximation algorithms use an
error budget to prune large tree fragments of this decision tree that only
account for a small probability mass. We introduce three approximation
approaches (\emph{eager}, \emph{lazy}, and \emph{hybrid}), each with a
different strategy for spending the error budget. The distributed
versions of these algorithms divide the exploration space into
fragments to be explored concurrently by distinct workers.

While the computation time can grow exponentially in the number of
input random variables in worst case, the structure of correlations can reduce it dramatically.
As shown experimentally, ENFrame's
algorithm for exact probability computation is orders of
magnitude faster than executing the user program in each possible
world.\smallskip

\noindent To sum up, the contributions of this paper are as follows:

\begin{itemize}[leftmargin=6mm]\vspace*{-2mm}\setlength{\itemsep}{-0.5mm}
\item We propose the ENFrame platform for processing probabilistic
  data.  ENFrame can evaluate user programs on probabilistic data with
  arbitrary correlations following the possible worlds
  semantics.

\item User programs are written in a fragment of Python that supports
  bounded-range loops, list comprehension, aggregates, and calls to
  external database engines. We illustrate ENFrame's features by giving programs for three clustering algorithms ($k$-means, $k$-medoids, and Markov
  clustering) and provide a formal specification of ENFrame's user language which can be used to write arbitrary programs for the platform.

\item User programs are annotated by ENFrame with events that are
  expressive enough to capture the correlations of the input, trace
  the program computation, and allow for probability
  computation.

\item ENFrame uses novel sequential and distributed algorithms for
  exact and approximate probability computation of event
  programs.

\item We implemented ENFrame's probability computation algorithms in
  C++.

\item We report on experiments with \kmedoids clustering of readings from partial discharge sensors in energy
  networks~\cite{Michel:2011}. We show orders-of-magnitude performance
  improvements of ENFrame's exact algorithm over the na\"ive approach of clustering in each
  possible world, of approximate over exact clustering, and of
  distributed over sequential algorithms.
\end{itemize}

The paper is organised as follows. Section~\ref{sec:user} introduces
the Python fragment supported by ENFrame along with encodings of
clustering algorithms. Section~\ref{sec:events} introduces our event
language and shows how user programs are annotated with events. Our
probability computation algorithms are introduced in
Section~\ref{sec:probcomp}, and experimentally evaluated in
Section~\ref{sec:experiments}. Section~\ref{sec:related} overviews
recent related work. \nop{Section~\ref{sec:conclusion} concludes the
  paper and gives a glimpse at future work.}

\section{ENFrame's User Language}
\label{sec:user}

\begin{figure*}
  \begin{minipage}[t]{0.55\textwidth}
    \begin{scriptsize}
      \begin{verbatim}
1:  (O, n) = loadData()        # list and number of objects
2:  (k, iter) = loadParams()   # number of clusters and iterations
3:  M = init()                 # initialise centroids

4:  for it in range(0,iter):   # clustering iterations
5:   InCl = [None] * k         # assignment phase
6:   for i in range(0,k):
7:    InCl[i] = [None] * n
8:    for l in range(0,n):
9:     InCl[i][l] = reduce_and(
10:       [dist(O[l],M[i]) <= dist(O[l],M[j]) for j in range(0,k)])
11:  InCl = breakTies2(InCl)   # each object is in exactly one cluster

12:  M = [None] * k            # update phase
13:  for i in range(0,k):
14:   M[i] = scalar_mult(invert(
15:     reduce_count([1 for l in range(0,n) if InCl[i][l]])),
16:     reduce_sum([O[l] for l in range(0,n) if InCl[i][l]]))
      \end{verbatim}
    \end{scriptsize}
  \end{minipage}
  \begin{minipage}[t]{0.45\textwidth}
    \begin{scriptsize}
      \hspace*{0em}$\forall i \text{ in } 0..n-1: O^i \equiv \Phi(o_i) \otimes \vec{o}_i$\\
      \hspace*{0em}$M^0_{-1} \equiv \Phi(o_{\pi(0)}) \otimes \vec{o}_{\pi(0)}; \ldots; M^{k-1}_{-1} \equiv \Phi(o_{\pi(k-1)}) \otimes \vec{o}_{\pi(k-1)}$\\

      \vspace*{0.8em}
      \hspace*{0em}$\forall \text{it} \text{ in } 0..\text{iter}-1:$\\
      \hspace*{0.7em}$\forall i \text{ in } 0..k-1:$\\
      \hspace*{1.4em}$\forall l \text{ in } 0..n-1:$\\
      \hspace*{2.1em}$\text{InCl}^{i,l}_{\text{it}} \equiv \bigwedge_{j=0}^{k-1} \bigl[
          \dist(O^l, M^i_{\text{it}-1}) \leq \dist(O^l, M^j_{\text{it}-1})\bigr]$\\

      \vspace*{2.1em}
      \hspace*{0.7em}\# Encoding of \texttt{breakTies2} omitted

      \vspace*{0.7em}
      \hspace*{0.7em}$\forall i \text{ in } 0..k-1:$\\
      \hspace*{1.4em}$M^i_{\text{it}} \equiv 
          \Bigl(\sum_{l=0}^{n-1} \text{InCl}^{i,l}_{\text{it}} \wedge \top\otimes 1 \Bigr)^{-1} \cdot
          \Bigl(\sum_{l=0}^{n-1} \text{InCl}^{i,l}_{\text{it}} \wedge O^l \Bigr)$
    \end{scriptsize}
  \end{minipage}%
  \vspace*{-5mm}%
  \caption{K-means clustering specified as user program (left) and simplified event program (right).}
  \label{fig:KMeansUserProgram}
\end{figure*}

\begin{figure*}
  \begin{minipage}[t]{0.55\textwidth}
    \begin{scriptsize}
      \begin{verbatim}
1:  (O, n, M) = loadData()     # M is a stochastic n*n matrix of
2:        # edge weights between the n nodes, O is list of nodes
3:  (r, iter) = loadParams()   # Hadamard power, number of iterations

4:  for it in range(0,iter):
5:   N = [None] * n            # expansion phase
6:   for i in range(0,n):
7:    N[i] = [None] * n
8:    for j in range(0,n):
9:     N[i][j] = reduce_sum([M[i][k]*M[k][j] for k in range(0,n)])

10:  M = [None] * n             # inflation phase
11:  for i in range(0,n):
12:   M[i] = [None] * n
13:   for j in range(0,n):
14:    M[i][j] = pow(N[i][j],r)*invert(
15:             reduce_sum([pow(N[i][k],r) for k in range(0,n)]))
      \end{verbatim}
    \end{scriptsize}
  \end{minipage}
  \begin{minipage}[t]{0.45\textwidth}
    \begin{scriptsize}
      \hspace*{0em}$\forall i \text{ in } 0..n-1: O^i \equiv \Phi(o_i) \otimes \vec{o}_i$\\

      \vspace*{1.8em}
      \hspace*{0em}$\forall \text{it} \text{ in } 0..\text{iter}-1:$\\
      \hspace*{0.7em}$\forall i \text{ in } 0..n-1:$\\
      \hspace*{1.4em}$\forall j \text{ in } 0..n-1:$\\
      \hspace*{2.1em}$N^{i, j}_{\text{it}} = \sum_{k=0}^{n-1} M^{i,k}_{\text{it-1}} \cdot M^{k, j}_{\text{it-1}}$ \\

      \vspace*{1.8em}
      \hspace*{0.7em}$\forall i \text{ in } 0..n-1:$\\
      \hspace*{1.4em}$\forall j \text{ in } 0..n-1:$\\
      \hspace*{2.1em}$M^{i, j}_{\text{it}} = \Bigl(\sum_{k=0}^{n-1}
          (N^{i,k}_{\text{it}})^r \Bigr)^{-1} \cdot (N^{i,j}_{\text{it}})^r$
    \end{scriptsize}
  \end{minipage}%
  \vspace*{-5mm}%
  \caption{Markov clustering specified as user program (left) and simplified event program (right).\label{fig:MclUserProgram}}%
  \vspace*{-5mm}%  
\end{figure*}

This section introduces the user language supported by ENFrame. Its
design is grounded in three main desiderata:

\begin{enumerate}[leftmargin=4mm]\setlength{\itemsep}{-1mm}\vspace*{-2mm}
\item It should naturally express common mining algorithms, allow to
  issue queries and manipulate their results.

\item User programs must be oblivious to the deterministic or
  probabilistic nature of the input data and to the probabilistic
  formalism considered.

\item It should be simple enough to allow for an intuitive and
  straightforward probabilistic interpretation.
\end{enumerate}
\vspace*{-.5em}

We settled on a subset of Python that can express, among others,
$k$-means, $k$-medoids, and Markov Clustering. In line with query
languages for probabilistic databases, where a Boolean query $Q$ is a
map $Q: D \to \{\text{true, false}\}$ for deterministic databases and
a Boolean random variable for probabilistic databases, every user
program has a sound semantics for both deterministic and probabilistic
input data: in the former case, the result of a clustering algorithm
is a deterministic clustering, in the latter case it is a probability
distribution over possible clusterings.

The user language comprises the following constructs:

\parhead{Variables and arrays.}  Variables can be of scalar types (real,
integer, or Boolean) or arrays. Examples of variable assignments:
\texttt{V = 2}, \texttt{W = V}, \texttt{M[2] = True}, or \texttt{M[i]
  = W}. Arrays must be initialised, e.g., for array \texttt{M} of
cardinality \texttt{k}: \texttt{M = [None] * k}.  Additionally, the
expression \texttt{range(0, n)} specifies the array
\texttt{[0,\dots,n-1]}.

\parhead{Functions.}  Scalar variables can be multiplied, exponentiated
(\texttt{pow(B, r)} for $B^r$), and inverted (\texttt{invert(B)} for
$1/B$). The function \texttt{dist(A,B)} is a distance measure on the
feature space between the vectors specified by arrays \texttt{A,B} of
reals; \texttt{scalar\_mult} is component-wise multiplication of an array
with a scalar.

\parhead{Reduce.}  Given a one-dimensional array \texttt{M} of some
scalar type, it can be \emph{reduced} to a scalar value by applying one of the functions
\texttt{reduce\_or}, \texttt{reduce\_sum}, \texttt{reduce\_count}. For
instance, for an array \texttt{B} of Booleans, the expression\break
\texttt{reduce\_and(B)} computes the Boolean conjunction of the truth
values in \texttt{B}, and the expression \texttt{reduce\_count(B)}
computes the number of elements in \texttt{B}.  For a two-dimensional
array of reals or integers, i.e., an array of vectors,
\texttt{reduce\_sum} computes the component-wise sum of the vectors.

\parhead{List comprehension.} Inside a \texttt{reduce}-function, anonymous arrays may be defined
using list comprehension. For example, given an array \texttt{B} of
Booleans of size \texttt{n}, the expression \texttt{reduce\_sum([1 for
i in range(0,n) if B[i]])} counts the number of \texttt{True} values
in \texttt{B}.

\parhead{Loops.} We only allow bounded-range loops; for any fixed
integer \texttt{n} and counter variable \texttt{i}, for-loops can be
defined by: \texttt{for i in range(0,n)}. This allows us to know the size of each constructed array at compile time.

\parhead{Input data.}  The special abstract primitive \texttt{loadData()}
is used to specify input data for algorithms. This function can be
implemented to statically specify the objects to be clustered, to load
them from disk, or to issue queries to a database. ENFrame supports
positive relational algebra queries with aggregates via the SPROUT
query engine for probabilistic data~\cite{FHO:VLDB12}.  The abstract
methods \texttt{loadParams()} and \texttt{init()} are used to set
algorithm parameters such as the number of iterations and clusters of a clustering algorithm.

\subsection{Clustering Algorithms in ENFrame}
We illustrate ENFrame's user language with three example data mining algorithms: $k$-means, $k$-medoids, and Markov Clustering. Figures~\ref{fig:KMedoidsUserProgram},~\ref{fig:KMeansUserProgram},~ and \ref{fig:MclUserProgram} list user
programs for these algorithms; we next discuss each of them.

\parhead{\emph{k}-means clustering.}  The $k$-means algorithm partitions a set
of $n$ data points $o_1,\ldots,o_n$ into $k$ groups of similar data
points. We initially choose a centroid $M^i$ for each cluster, i.e., a
data point representing the cluster centre (initialisation phase). In
successive iterations, each data point is assigned to the cluster with
the closest centroid (assignment phase), after which the centroid is recomputed for each cluster (update phase). The algorithm
terminates after a given number of iterations or after reaching
convergence. Note that our user language does not support fixpoint
computation, and hence checking convergence.

Figure~\ref{fig:KMeansUserProgram} implements $k$-means. The set \texttt{O} of
$n$ input objects is retrieved using a \texttt{loadData} call. Each object is
represented by a feature vector (i.e., array) of reals. We then load the
parameters \texttt{k}, the number \texttt{iter} of iterations, and initialise
cluster centroids \texttt{M} (line 3).  The initialisation phase has a
significant influence on the clustering outcome and convergence. We assume that initial centroids have been chosen, for example by using a heuristic~\cite{Omurca:2011:IFART}. Subsequently, an array \texttt{InCl} of Booleans is computed such that \texttt{InCl[i][l]} is
\texttt{True} if and only if \texttt{M[i]} is the closest centroid to object
\texttt{O[l]} (lines 5--10); every object is then assigned to its closest
cluster. Since two clusters may be equidistant to an object, ties are broken
using the \texttt{breakTies2} call (line 11); it fixes an order of the clusters
and enforces that each object is only assigned to the \emph{first} of its
potentially multiple closest clusters. Next, the new cluster centroids
\texttt{M[i]} are computed as the centroids of each cluster (lines 12--16). The
assignment and update phases are repeated \texttt{iter} times (line 4).

\parhead{\emph{k}-medoids clustering.}  The $k$-medoids algorithm is almost identical to $k$-means, but elects $k$ cluster \emph{medoids} rather than centroids: these are cluster members that minimise the sum of distances to all other objects in the cluster.  The assignment phase is the
same as for $k$-means (lines 5--11), while the update phase is more
involved: We first compute an array \texttt{DistSum} of sums of
distances between each cluster medoid and all other objects in its
cluster (lines 12--17), then find one object in each cluster that
minimises this sum (lines 18--24), and finally elect these objects as
the new cluster medoids \texttt{M} (lines 25--27).  The last step uses
\texttt{reduce\_sum} to select exactly one of the objects in a cluster
as the new medoid, since for each fixed \texttt{i} only one value in
\texttt{Centre[i][l]} is \texttt{True} due to the tie-breaker in line
24.

\parhead{Markov clustering (MCL).} MCL is a fast and scalable unsupervised 
cluster algorithm for graphs based on simulation of stochastic flow in
graphs~\cite{vanDongen:PhD:2000}. Natural clusters in a graph are
characterised by the presence of many edges within a cluster and few
edges across clusters. MCL simulates random walks within a graph by
alternating two operations: \emph{expansion} and
\emph{inflation}. Expansion corresponds to computing random walks of
higher length. It associates new probabilities with all pairs of
nodes, where one node is the point of departure and the other is the
destination. Since higher length paths are more common within clusters
than between different clusters, the probabilities associated with
node pairs lying in the same cluster will, in general, be relatively
large as there are many ways of going from one to the other. Inflation
has the effect of boosting the probabilities of intra-cluster walks
and demoting inter-cluster walks. This is achieved without a priori
knowledge of cluster structure; it is the result of cluster structure
being present.

Figure~\ref{fig:MclUserProgram} gives the MCL user program. Expansion
coincides with taking the power of a stochastic matrix \texttt{M} using the
normal matrix product (i.e. matrix squaring). Inflation corresponds
to taking the Hadamard power of a matrix (taking powers entry-wise). It is
followed by a scaling step to maintain the stochastic property, i.e. the matrix elements correspond to probabilities that sum up to 1 in each column.

Section~\ref{sec:events} discusses the probabilistic interpretation of
the computation of the above three clustering algorithms.

%%%%%%%%%%%%%%%%%
\begin{figure}
  \hspace*{-3mm}\begin{minipage}{\linewidth}\begin{small}\vspace*{-2mm}
  \begin{align*}
    \text{LOOP}  &::= \text{\{ \{DECL\}} \text{\{ for ID in RANGE:} \text{ \{LOOP\} \} \}} \\
    \text{DECL}   &::= \text{(ID = EXPR) | (`('\{ID, \} ID `)' = EXT) } \\
    \text{EXPR}   &::= \text{LIT | ID | [None] `*' EXPR) | } \text{(EXPR COMP EXPR) |} \\
    &    \text{(REDUCE `(' LCOMPR `)') | } \text{(pow`('EXPR, EXPR`)') | }\\
    &    \text{(invert`('EXPR`)') | } \text{(EXPR `*' EXPR) | (EXPR `+' EXPR) |} \\
    &    \text{(scalar\_mult`('EXPR, EXPR`)') | } \text{(breakTies`('EXPR`)')}\\
    \text{LCOMPR} &::= \text{[EXPR for ID in RANGE if EXPR]} \\
    \text{REDUCE} &::= \text{reduce\_and | reduce\_or | reduce\_sum |} \\
    &    \text{reduce\_mult | reduce\_count} \\
    \text{RANGE}  &::= \text{range(EXPR, EXPR)} \\
    \text{COMP}   &::= \text{`<' | `>' | `==' | `<=' | `>='} \\
    \text{EXT}    &::= \text{loadData() | loadParams() | init()} \\
    \text{ID}     &::= \text{An identifier} \\
    \text{LIT}    &::= \text{A (Boolean, integer, float) literal}
  \end{align*}
  \end{small}\end{minipage}%
  \vspace*{-2mm}%
  \caption{The grammar of the user language.\label{fig:UserLanguage}}%
\end{figure}
%%%%%%%%%%%%%%%%%

\subsection{Syntax of the User Language}
Figure~\ref{fig:UserLanguage} specifies the formal grammar for the
language of user programs. A program is a sequence of
declarations (DECL) and loop blocks (LOOP), each of which may again
contain declarations and nested loops.  The language allows to assign
expressions (EXPR) to variable identifiers (ID).  An expression may be
a Boolean, integer, or float constant (LIT), an identifier, an array
declaration, the result of a Boolean comparison between expressions,
or the result of such operations as sum, product, inversion, or
exponentiation. The result of a reduce operation on an anonymous array
created through list comprehension (LCOMPR), and the result of
breaking ties in a Boolean array give rise to expressions; we
elaborate on these two constructions below.

In addition to the syntactic structure as defined by the grammar, programs
have to satisfy the following constraints:

\parhead{Bounded-range loops.}  The parameters to the range construct
must be integer constants (or immutable integer-valued variables).
This restriction ensures that for-loops (LOOP) and list comprehensions
(LCOMPR) are of bounded size that is known at compile time.

\parhead{Anonymous arrays via list comprehension.}  List comprehension
may only be used to construct one-dimensional arrays of base types,
i.e., arrays of integers, floats, or Booleans.

\parhead{Breaking ties.}  Clustering algorithms require explicit handling
of ties: For instance, if two objects are equidistant to two distinct
cluster centroids in $k$-means, the algorithm has to decide which
cluster the object will be assigned to. In ENFrame programs, the
membership of objects to clusters can be encoded by a Boolean array
like \texttt{InCl} such that \texttt{InCl[i][l]} is true if and only
if object \texttt{l} is in cluster \texttt{i}. In this context, a tie
is a configuration of \texttt{InCl} in which for a fixed object
\texttt{l}, \texttt{InCl[i][l]} is \texttt{True} for more than one
cluster \texttt{i}. We explicitly break such ties using the function
\texttt{breakTies2(M)}. For each fixed value \texttt{i} of the second
dimension (hence the 2 in the function name) of the 2-dimensional
array \texttt{M}, it iterates over the first dimension of \texttt{M}
and sets all but the first \texttt{True} value of \texttt{M[i][l]} to
\texttt{False}. Symmetrically, the function \texttt{breakTies1(M)}
fixes the first dimension and breaks ties in the second dimension of
\texttt{M}, and \texttt{breakTies(M)} breaks ties in a one-dimensional
array.

\section{Tracing Computation by Events}
\label{sec:events}

The central concept for representing user programs in ENFrame is that
of \emph{events}. Each event is a concise syntactic encoding of a
random variable and its probability distribution. This section
describes the syntax and semantics of events and event programs, and
finally explains how ENFrame programs written in the user language
from Section~\ref{sec:user} can be translated to event programs.

The key features of events and event programs are:
\begin{itemize}\setlength{\itemsep}{1mm}\vspace*{-0.5em}
\item Events can encode arbitrarily correlated, discrete probability
  distributions over input objects. In particular, they can succinctly
  encode instances of such formalisms as Bayesian networks and
  pc-tables. The input objects and their correlations can be
  explicitly provided, or imported via a positive relational algebra
  query with aggregates over
  pc-tables~\cite{FHO:VLDB12}.\vspace*{-0.5em}

\item By allowing non-Boolean events, our encoding is exponentially
    more succinct than an equivalent purely Boolean
    description.\vspace*{-0.5em}

\item Each event has a well-defined probabilistic semantics that
    allows to interpret it as a random variable.\vspace*{-0.5em}

\item The iterative nature of many clustering algorithms carries over to event
    programs, in which events can be defined by means of nested loops. This
    construction together with the ability to reuse existing, named events in
    the definition of new, more complex events leads to a concise encoding of a
    large number of distinct events.
\end{itemize}

\begin{example}\textbf{Clustering in possible worlds.}
  We start by presenting an instructive example of $k$-medoids clustering under
  possible worlds semantics. Let $o_0, \dots, o_3$ be objects in the feature
  space as shown below. They can be clustered into two clusters using $k$-medoids
  with medoids $o_1$ and $o_3$.
  
  \vspace*{-2.5mm}\begin{center}\begin{small}\scalebox{0.7}{\begin{tikzpicture}
          \draw [rounded corners=5pt,draw=black!70,fill=red!20,densely dashed] (-0.45,0.5) rectangle  +(5.9,1);
          \draw [rounded corners=5pt,draw=black!70,fill=blue!20,densely dashed] (8.55,0.5) rectangle  +(0.9,1);
          \draw[draw=black!50] (-0.6,0.4)  grid (9.6,1.6);
          \node (1) at (0, 1) [circle,draw=black!70,fill=red!50] {$o_0$};
          \node (2) at (2, 1) [circle,draw=black!70,fill=green!50,very thick] {$o_1$};
          \node (3) at (5, 1) [circle,draw=black!70,fill=blue!50] {$o_2$};
          \node (4) at (9, 1) [circle,draw=black!70,fill=yellow!50,very thick] {$o_3$};
  \end{tikzpicture}}\end{small}\end{center}\vspace*{-2.5mm}

  In order to go from deterministic to uncertain objects, we associate each object
  $o_l$ with a Boolean propositional formula $\Phi(o_l)$ -- the event -- over a
  set of independent Boolean random variables $\bfX$. The possible valuations
  $\nu: \bfX \to \{\text{\true, \false}\}$ define the the \emph{possible worlds} of the input objects: for each valuation $\nu$ there exists on world that contains exactly those objects $o_l$ for which $\Phi(o_l)$ is \true under $\nu$. The probability of a world is the product of the
  probabilities of the variables $x \in \bfX$ taking a truth value $\nu(x)$.

  Let us assume that the objects have the following events:
  
  \begin{small}$\Phi(o_0) = x_1 \vee x_3,~\Phi(o_1) = x_2,~\Phi(o_2) =
    x_3,~\Phi(o_3) = \neg x_2 \wedge x_4$.\end{small}

  Distinct worlds can have different clustering results, as
  exemplified next. The world defined by
  $\{x_1\mapsto\top,x_2\mapsto\bot,x_3\mapsto\top,x_4\mapsto\top\}$
  consists of objects $o_0$, $o_2$, and $o_3$, for which $k$-medoids
  clustering yields:

  \vspace*{-2.5mm}\begin{small}\begin{center}\scalebox{0.7}{\begin{tikzpicture}
          \draw [rounded corners=5pt,draw=black!70,fill=red!20,densely dashed] (-0.45,0.5) rectangle  +(0.9,1);
          \draw [rounded corners=5pt,draw=black!70,fill=blue!20,densely dashed] (4.55,0.5) rectangle  +(4.9,1);
          \draw[draw=black!50] (-0.6,0.4)  grid (9.6,1.6);
          \node (1) at (0, 1) [circle,draw=black!70,fill=red!50,very thick] {$o_0$};
          \node[black!30] (2) at (2, 1) [circle,draw=black!20,fill=green!10,densely dotted,thick] {$o_1$};
          \node (3) at (5, 1) [circle,draw=black!70,fill=blue!50,very thick] {$o_2$};
          \node (4) at (9, 1) [circle,draw=black!70,fill=yellow!50] {$o_3$};
  \end{tikzpicture}}\end{center}\end{small}\vspace*{-2.5mm}

  Similarly, the worlds defined by
  $\{x_1\mapsto\top,x_2\mapsto\top,x_3\mapsto\top\}$ and any assignment
  for $x_4$, yields:
  
  \vspace*{-2.5mm}\begin{small}\begin{center}\scalebox{0.7}{\begin{tikzpicture}
          \draw [rounded corners=5pt,draw=black!70,fill=red!20,densely dashed] (-0.45,0.5) rectangle  +(2.9,1);
          \draw [rounded corners=5pt,draw=black!70,fill=blue!20,densely dashed] (4.55,0.5) rectangle  +(0.9,1);
          \draw[draw=black!50] (-0.6,0.4)  grid (9.6,1.6);	
          \node (1) at (0, 1) [circle,draw=black!70,fill=red!50,very thick] {$o_0$};
          \node (2) at (2, 1) [circle,draw=black!70,fill=green!50] {$o_1$};
          \node (3) at (5, 1) [circle,draw=black!70,fill=blue!50,very thick] {$o_2$};
          \node[black!30] (4) at (9, 1) [circle,draw=black!20,fill=yellow!10,densely dotted,thick] {$o_3$};
  \end{tikzpicture}}\end{center}\end{small}\vspace*{-2.5mm}%

  The probability of a query ``Are $o_1$ and $o_2$ in the same cluster?'' is the
  sum of the worlds in which $o_1$ and $o_2$ are in the same cluster. 
  \punto
  \label{ex:ClusteringInWorlds}
\end{example}

Events do not only encode the correlations and probabilities of input objects,
but can symbolically encode the entire clustering process. We illustrate this in
the next example.

\begin{example}\textbf{Symbolic encoding of $k$-means by events.}
  We again assume four input objects $o_0$, \dots, $o_3$ with their
  respective events $\Phi(o_l)$. This example introduces
  \emph{conditional values} (c-values) which are expressions of the
  form $\Phi \otimes v$, where $\Phi$ is a Boolean formula and $v$ is a
  vector from the feature space. Intuitively, this c-value takes the
  value $v$ whenever $\Phi$ evaluates to \true, and a special \emph{undefined}
  value when $\Phi$ is \false. C-values can be added and multiplied;
  for example, the expression $\Phi\otimes v + \Psi\otimes w$
  evaluates to $v+w$ if $\Phi$ and $\Psi$ are \true, or to $v$ if
  $\Phi$ is \true and $\Psi$ is \false, etc.
  
  Equipped with c-values, an initialisation of $k$-means with $k=2$
  can for instance be written in terms of two expressions $M^0 =
  \Phi(o_0) \otimes o_0 + \neg\Phi(o_0) \otimes o_2$ and $M^1 = \top
  \otimes 0.5\cdot (o_1 + o_3)$: Centroid $M^0$ is set to object $o_0$
  if $\Phi(o_0)$ is \true and to $o_2$ if $\Phi(o_0)$ is \false;
  centroid $M^1$ is always set to the geometric centre of $o_1$ and
  $o_3$.

  In the assignment phase, each object is assigned to its nearest cluster centroid. The condition InCl$^{i,l}$ for object $o_l$ being
  closest to $M^i$ can be written as the Boolean event  
  \noindent\begin{minipage}{\columnwidth}\vspace*{1mm}\noindent\begin{math}
  \text{InCl}^{i,l} \equiv \bigwedge_{j=0}^{1} \bigl[\dist(\Phi(o_l)
  \otimes o_l, M^i) \leq \dist(\Phi(o_l) \otimes o_l, M^j)\bigr]
  \end{math},\vspace*{1mm}\end{minipage}
  
  \noindent which encodes that the distance from $o_l$ to centroid $M^0$ is
  smaller than the distance to centroid $M^1$.

  Given the Boolean events InCl$^{i,l}$, we can represent the centroid of cluster
  $i$ for the next iteration by the expression
  \noindent\begin{minipage}{\columnwidth}\centering\noindent\begin{math}
  \Bigl(\sum_{l=0}^{3} \text{InCl}^{i,l} \otimes 1 \Bigr)^{-1} \cdot
    \Bigl(\sum_{l=0}^{3} \text{InCl}^{i,l} \otimes o_l \Bigr)
  \end{math},\vspace*{1mm}\end{minipage}
  which specifies a random variable over possible cluster centroids conditioned
  on the assignments of objects to clusters as encoded by InCl$^{i,l}$. This
  expression is exponentially more succinct than an equivalent purely Boolean
  encoding of cluster centroids, since the later would require one Boolean
  expression for each subset of the four input objects.
  \punto
  \label{ex:KMeansSymbolic}
\end{example}

The event programs corresponding to the three user programs for $k$-means,
$k$-medoids, and MCL are given on the right side of
Figures~\ref{fig:KMedoidsUserProgram}--\ref{fig:MclUserProgram}. In addition to the constructs
introduced in Example~\ref{ex:KMeansSymbolic}, they use event declarations that
assign identifiers to event expressions, and $\forall i$-loops that specify
sets of events parametrised by $i$. The remainder of this section specifies
the formal syntax and semantics of event programs, and gives a translation from 
user to event programs.

\subsection{Syntax of Event Expressions}
The grammar for event expressions is as follows:

\begin{small}\vspace*{-4mm}
  \begin{align*}
    \text{VAL}   &::= \text{A scalar or feature vector} \\
    \text{INT}   &::= \text{Any integer}\\
    \text{CVAL}  &::= \text{EVENT$\otimes$VAL
                          | CVAL$^{-1}$ 
                          | CVAL+CVAL
                          | CVAL$^\text{INT}$
                          |
                      } \\
                      & \text{CVAL $\cdot$ CVAL | \text{dist}(CVAL, CVAL) | \text{EVENT $\wedge$ CVAL}} \\
    \text{COMP}  &::= \text{$\leq | \geq | = | < | >$} \\
    \text{ATOM}  &::= \text{[CVAL COMP CVAL]} \\
    \text{EID}   &::= \text{Elements of a set of event identifiers} \\
    \text{EVENT} &::= \text{Propositional formula over $\bfX$, EID, ATOM}
  \end{align*}
\end{small}\vspace*{-4mm}

The main constructs are:

\parhead{Conditional values.}
Reals and feature vectors are denoted by VAL. Together with a propositional
formula, they give rise to a \emph{conditional value} (CVAL), c-value for short. 

\parhead{Functions of conditional values.}
Very much like scalars and feature vectors, c-values can be added,
multiplied, and exponentiated. Additionally, the distance between two c-values
yields another c-value. In addition to the binary operations specified in the
grammar (e.g., CVAL+CVAL), we allow $\sum$- and $\prod$-expressions (see
Figure~\ref{fig:KMeansUserProgram}).

\parhead{Event expressions.}
Event expressions (EVENT) are propositional formulas over constants
$\top$ (\true), $\bot$ (\false), a set $\bfX$ of Boolean random variables,
event identifiers, and propositions defined by ATOM: [CVAL COMP CVAL]
represents the truth value obtained by comparing two c-values.

\subsection{Semantics of Event Expressions}
The semantics of event expressions is defined by extending a Boolean valuation
$\nu: \bfX \to \{\text{\true, \false}\}$ to a valuation of c-values and event
expressions. We define in the sequel how $\nu$ acts on each of the expression
types in the grammar.
The base cases of this mapping are the standard algebraic operations on scalars
and the feature space, extended by special \emph{undefined} elements as follows.

We extend the reals (and their operations $+$, $\cdot$, $()^{-1}$) by a special
element $u$ (for \emph{undefined}) such that $0^{-1} = u$. Operators $+, \cdot$
propagate $u$ as $u+x = x$ and $u \cdot x = u$ for any real $x$. For any other
reals $x,y$, $+$ and $\cdot$ are as usual.  For example, $5 \cdot (3 - 3)^{-1} =
5 \cdot u = u$. 

Similarly, we extend the feature space by an element $\bfu$. For any real $a$
and feature vector, $u$ and $\bfu$ are propagated as $u \cdot \bfx = \bfu$,
$\bfu + \bfx = \bfx$, $a \cdot \bfu = \bfu$, and $\bfu \cdot \bfx = u$.

The grammar for event programs does not distinguish between
scalars and feature vectors for the sake of notational clarity. The following
description implicitly assumes that the expressions are well-typed; 
e.g., the expression $\text{dist}(x, y)$ is only defined for vector-valued
variable symbols $x,y$.

\parhead{CVAL.}
Conditional values of the form EVENT$\otimes$VAL have an if-then-else semantics: If EVENT 
evaluates to \true, then EVENT$\otimes$VAL evaluates to VAL, else it evaluates to
$u$ (or $\bfu$ for vector-valued c-values); the recursively constructed CVAL
expressions have the natural recursive semantics that ultimately defaults to $+$
and $\cdot$ for scalars and feature vectors.

\begin{small}\vspace*{-4mm}\begin{align*}
  \nu(\text{EVENT}\otimes\text{VAL})      &=
      \begin{cases}
        \text{VAL}, &\text{if } \nu(\text{EVENT}) = \text{\true} \\
        u \text{ ($\bfu$, resp.)} &\text{otherwise}
      \end{cases} \\
  \nu(\text{CVAL}_1 + \text{CVAL}_2)     &= \nu(\text{CVAL}_1) + \nu(\text{CVAL}_2) \\
  \nu(\text{CVAL}_1 \cdot \text{CVAL}_2) &= \nu(\text{CVAL}_1) \cdot \nu(\text{CVAL}_2) \\
  \nu(\text{CVAL}^{-1})                  &= \nu(\text{CVAL})^{-1} \\
  \nu(\text{\text{dist}(CVAL$_1$, CVAL$_2$)})      &=
      \begin{cases}
        u, \;\text{if } \nu(\text{CVAL}_1) = u \text{ or } \\
        \hspace*{4.2mm} \nu(\text{CVAL}_2) = u  \\
        \text{\text{dist}($\nu$(CVAL$_1$), $\nu$(CVAL$_2$))}, \;\text{else}
      \end{cases} \\
  \nu(\text{CVAL}^{\text{INT}})          &= \nu(\text{CVAL})^{\text{INT}}\\
  \nu(\text{EVENT $\wedge$ CVAL})     &= 
      \begin{cases}
        \nu(\text{CVAL}), &\text{if } \nu(\text{EVENT}) = \text{\true} \\
        u \text{ ($\bfu$, resp.)} &\text{otherwise}
      \end{cases} \\
\end{align*}\vspace*{-7mm}\end{small}
\vspace*{-3mm}

\parhead{ATOM, EVENT.}
Comparisons $\nu(\text{[CVAL$_1$ $\theta$ CVAL$_2$]})$ for $\theta \in \{\leq, \geq, =, <, >\}$ between two c-values
evaluate to \false if they are both defined ($\nu(\text{CVAL$_1$}) \neq u$ and
$\nu(\text{CVAL$_2$}) \neq u$)
and the comparison does not hold; otherwise (i.e.~if at least one of the
c-values is undefined, or if the comparison holds), it evaluates to \true.
% For every comparison $\theta \in \{\leq, \geq, =, <, >\}$ we obtain:
%   \begin{align*}
%     \nu(\text{[CVAL$_1$ $\theta$ CVAL$_2$]})      &=
%       \begin{cases}
%         \text{false}, &\text{if } \nu(\text{CVAL}_1) \neq u \text{ and } \\
%                       &\nu(\text{CVAL}_2) \neq u \text{ and not} \\
%                       &\nu(\text{CVAL}_1) \;\theta\; \nu(\text{CVAL}_2)\\
%         \text{true}, &\text{else}
%       \end{cases} \\
%   \end{align*}
The semantics of the Boolean propositional EVENT expressions is standard,
i.e.~by propagating $\nu$ through the propositional operators $\wedge, \vee,
\neg$.  For instance $\nu(\text{EVENT}_1 \wedge \text{EVENT}_2)$ evaluates to
\true if $\nu(\text{EVENT}_1) = \nu(\text{EVENT}_2) = \text{\true}$, and to \false
otherwise.

\subsection{Probabilistic Semantics of Events}
We next give a probabilistic interpretation of event expressions that explains
how they can be understood as random variables: Boolean event expressions
(EVENT) give rise to Boolean random variables, and conditional values (CVAL) give rise
to random variables over their respective domain.

For every random variable $x \in \bfX$, we denote by $P_x[\text{\true}]$ and
$P_x[\text{\false}]$ the probability that $x$ is \true or \false, respectively; we
also simply write $P_x$ for $P_x[\text{\true}]$. \nop{Wlog, we assume that $0 <
  P_x < 1$.}  Let $\Omega = \{ \nu: \bfX \to \{\text{\true, \false}\}\}$ be the set
of mappings from the random variables $\bfX$ to \true and \false.

\begin{definition}[Induced Probability Space]
  Together, the probability mass function
    $\Pr(\nu) = \prod_{x\in\bfX} P_x[\nu(x)]$
  for every sample $\nu\in\Omega$, and the probability measure
    $\Pr(E) = \sum_{\nu\in E} \Pr(\nu)$ for $E\subseteq \Omega$
  define a probability space $(\Omega, 2^\Omega, \Pr)$ that we call the
  \emph{probability space induced by} $\bfX$.
\end{definition}

An event expression $E$ is a random variable over the probability space induced by $\bfX$ with probability distribution
\vspace*{-1mm}\begin{equation*}
  P_E[s] = \Pr\bigl( \{ \nu\in\Omega \;|\; \nu(E)\mathord{=}s \} \bigr) = \sum_{\substack{\nu\in\Omega: \nu(E)\mathord{=}s}} \Pr(\nu).
  \label{eq:ProbDistributionEventExpression}
\end{equation*}\vspace*{-2mm}

By virtue of this definition, every Boolean event expression becomes a Boolean
random variable, and real-valued (vector-valued) c-values become random
variables over the reals (the feature space).

\subsection{Event Programs}
Event programs are imperative specifications that define a finite set of named
c-values and event expressions. The grammar for event programs is as follows:

\begin{small}\vspace*{-4mm}
  \begin{align*}
    \text{INT} &::= \text{A positive integer} \\
    \text{VAR} &::= \text{A variable symbol} \\
    \text{LOOP}  &::= \text{\{ \{DECL\} \{ $\forall$ VAR in INT..INT: \{LOOP\} \} \} } \\
    \text{DECL}  &::= \text{EID $\equiv$ EVENT}
  \end{align*}
\end{small}\vspace*{-4mm}

Event programs consist of a sequence of event declarations
(DECL) and nested loops (LOOP) of event declarations. 

A central concept is that of event identifiers (EID); it is required that event
declarations are immutable, i.e.~each distinct EID may only be assigned once to
an event expression. Inside a $\forall i$-loop, identifiers can be parametrised
by $i$ to create a distinct identifier in each iteration of the loop.

The meaning of an event program is simply the set of all named and
\emph{grounded} c-value and event expressions defined by the program; grounded
here means that all identifiers in expressions are recursively resolved and
replaced by the referenced expressions. For declarations outside of loops, this
is clear; each declaration inside of (nested loops) is instantiated for each
value of the loop counter variables.

\subsection{From User Programs to Event Programs} 
The translation of user to event programs has two main challenges: (i)
Translating mutable variables and arrays to immutable events, and (ii)
translating function calls such as \texttt{reduce\_*}. We cover these two
issues separately.

\parhead{From mutable variables to immutable events.}
It is natural to reassign variables in user language programs, for example when
updating $k$-means centroids in each iteration based on the cluster configuration
of the previous iteration. In contrast, events in event programs are
\emph{immutable}, i.e., can be assigned only once. The translation from the user
language
to the event language utilises a function \texttt{getLabel} that generates for each
user language variable $M$ a sequence of unique event identifiers whose
lexicographic order reflects the sequence of assignments of $M$.

The basic idea of \texttt{getLabel} is to first identify the nested loop blocks
of the given user language program, and then to establish a counter for each distinct
variable symbol $M$ and each block. An assignment of a variable within $k$
nested blocks corresponds to an event identifier of the form
$M_{c_1.\;...\;.c_k}$ where $c_1, \dots, c_k$ are the $k$ counters for the $k$
blocks. Within each block, its corresponding counter is incremented for every
assignment of its variable symbol. When going from one block into a nested inner
block, the counters for the outer blocks are kept constant while the counter for
the inner block is incremented as $M$ is reassigned in the inner block.

Special attention must be paid to the encoding of entering and leaving a block:
In order to carry over the reference to a variable $M_{c_1.\;...\;.c_k}$ to
the next block at level $k+1$, we establish a copy $M_{c_1.\;...\;.c_k.(-1)}
\equiv M_{c_1.\;...\;.c_k}$, such that the first access to $M$ in the block
may access its last assignment of $M$ via $M_{c_1.\;...\;.c_k.(-1)}$.
Similarly, the last assignment of a variable in the inner block is passed back
to the outer block by copying the last identifier of an inner block to the next
identifier of the outer block.

\begin{example}
  Consider the following user language program (left) and its
  translation to an event program (right).

  \smallskip
  \hspace*{-2em}
  \begin{small}
    \begin{tabular}{l@{\;}l@{\quad\;}l@{\;}l}
      \texttt{1:} & \verb|M = 7| &                     A:& \hspace*{0em}$M_{0} \equiv 7$ \\
      \texttt{2:} & \verb|M = M+2| &                   B:& \hspace*{0em}$M_{1} \equiv M_0 + 2$ \\
      \texttt{3:} & \verb|for i in range(0,2):| &      C:& \hspace*{0em}$M_{1.-1} \equiv M_1$ \\
                  & &                                  D:& \hspace*{0em}$\forall i \text{ in } 0..1:$ \\
      \texttt{4:} & \verb| M = M+i| &                  E:& \hspace*{0.7em}$M_{1.(2i)} \equiv M_{1.(2i-1)} + i$ \\
      \texttt{5:} & \verb| for j in range(0,3):| &     F:& \hspace*{0.7em}$M_{1.(2i).-1} \equiv M_{1.(2i)}$ \\
                  & &                                  G:& \hspace*{0.7em}$\forall j \text{ in } 0..2:$ \\
      \texttt{6:} & \verb|  M = M+1| &                 H:& \hspace*{1.4em}$M_{1.(2i).j} \equiv M_{1.(2i).(j-1)} + 1$ \\
                  & &                                  I:& \hspace*{0.7em}$M_{1.(2i+1)} \equiv M_{1.(2i).2}$\\
                  & &                                  J:& \hspace*{0em}$M_{2} \equiv M_{1.(2\cdot 1 + 1)}$ \\
      \texttt{7:} & \verb|M = M+1| &                   K:& \hspace*{0em}$M_{3} \equiv M_{2} + 1$
  \end{tabular}
  \end{small}

  The user language program has three nested blocks. Within each
  block, the respective counter is incremented for each assignment of $M$:
  $M_0, \dots, M_3$ for the outer block, $M_{1.0}, \dots, M_{1.3}$ in the second 
  block, and $M_{1.(2i).0}, \dots, M_{1.(2i).2}$ for the innermost block.
  The encodings for entering and leaving a block are in lines C and F, and 
  lines I and J, respectively.
  \punto
\end{example}

\parhead{Translation of arrays.}
Since arrays in a user language program have a known fixed size, their translation
is straightforward: A $k$-dimensional array $M[n_1]...[n_k]$ translates
to $\prod_i n_i$ distinct identifiers $M^{0, \dots, 0}, \dots, M^{n_1-1,
  \dots, n_k-1}$.

\parhead{Translation of \texttt{reduce\_*} calls.}
According to the grammar in Figure~\ref{fig:UserLanguage}, reduce-operations can
only be applied to anonymous arrays created by list comprehension. The
expression \texttt{reduce\_and([EXPR for ID in range(FROM, TO) if COND]} is
translated to the Boolean event $\bigwedge_{\text{ID}=\text{FROM}}^{\text{TO}-1}
\text{COND} \wedge \text{EXPR}$.  Symmetrically, \texttt{reduce\_or} translates
to $\bigvee$, \texttt{reduce\_sum} to $\sum$, and \texttt{reduce\_mult} to
$\prod$. A call to \texttt{reduce\_count([EXPR for ID in range(FROM, TO) if
  COND])} translates to the event
$\sum_{\text{ID}=\text{FROM}}^{\text{TO}-1} \text{COND} \otimes 1$.

\section{Probability Computation}
\label{sec:probcomp}

The probability computation problem is known to be \#P-hard already
for simple events representing propositional formulas such as positive
bipartite formulas in disjunctive normal form~\cite{Ball83}. In
ENFrame, we need to compute probabilities of a large number of
interconnected complex events. Although the worst-case complexity
remains hard, we attack the problem with three complementary
techniques: (1) bulk-compile all events into one decision tree while
exploiting the structure of the events to obtain smaller trees, (2)
employ approximation techniques to prune significant parts of the decision tree, and ultimately (3) distribute the
compilation by assigning distinct distributed workers to explore disjoint parts of
the tree.

We next introduce the bulk-compilation technique, look at three
approximation approaches, and discuss how to distribute the probability computation.

\subsection{Compilation of event programs}

\begin{algorithm}[t]\begin{scriptsize}\SetInd{1.6mm}{1.6mm}
%todo:
% - make 'targets reached' and 'targets approximated' more formal
% - 
\DontPrintSemicolon
\Comment{{\color{blue}Blue} comments and pseudocode are related to $\varepsilon$-approx.}
\FuncSty{Compile}(\DataSty{network $D$, {\color{blue}absolute error $\varepsilon$}})
\Begin{
 \Comment{Initialise initial (empty) masks for nodes in the network}
 \lForEach{$v_i \in D$}{$M[v_i] \gets \text{unknown}$}\\
 
 \ForEach{$t_i \in \text{targets}(D)$}{
  $t_i.\emphtext{problower} \gets 0$\Comment*[f]{initial probability lower bound: 0}\\
  $t_i.\emphtext{probupper} \gets 1$\Comment*[f]{initial probability lower bound: 1}\\
  {\color{blue}$E[t_i] \gets 2\varepsilon$\Comment*{error budget (for exact, $\varepsilon = 0$)}}
 }
 \vspace*{0.1cm}
 \FuncSty{dfs}($D,M,\{\,\},{\color{blue}E}$) \Comment*{empty DFS branch $\nu = \{\,\}$, $\Pr(\nu) = 1$}
}
\vspace*{0.2cm}
\FuncSty{dfs}(\DataSty{network $D$, masks $M$, branch $\nu$, error budgets $E$})
\Begin{
 {\color{blue}
 \If(\Comment*[f]{sufficient budget}){$\forall t_i \in \text{targets}(D): E[t_i] \geq \Pr(\nu)$}{
  \lForEach{$t_i \in T$}{$E'[t_i] \gets E[t_i] - \Pr(\nu)$}\\
  \Return{$E'$}
 }}
 \If(\Comment*[f]{propagate variable mask into DAG}){$\nu \neq \varnothing$}{
  $M[\emphtext{top}(\nu).\emphtext{var}] \gets \emphtext{top}(\nu).\emphtext{val}$\\
  $M \gets$ \FuncSty{mask}$\left(D, M, \emphtext{top}(\nu).\emphtext{var}, \text{\textsc{null}}, \Pr(\nu)\right)$
 }
 \lIf{$\forall t_i \in \text{targets}(D):~(t_i.\text{probupper} - t_i.\text{problower} \leq {\color{blue}2\varepsilon}$ \textbf{or}\\
    $~~~~M[t_i] \neq \emphtext{unknown})$}{
  \Return{{\color{blue}$E$}}\Comment*{all reached/approx.}
 }
 \vspace*{0.1cm}
 $x \gets \emphtext{nextVariable}(\nu)$\\
 \vspace*{0.1cm}
 
 {\color{blue}\Comment{error budget for left DFS-branch}\vspace*{-1mm}
 \lForEach{$t_i \in \text{targets}(D)$}{$E_{\text{left}}[t_i] \gets \frac{E[t_i]}{2}$}\\
 \vspace*{0.1cm}}
 
 \Comment{DFS left branch, storing the {\color{blue}residual error budget $E_\text{left}'$}}\vspace*{-0.5mm}
 ${\color{blue}E_\emphtext{left}'} \gets \FuncSty{dfs}\left(D, M_\text{left}, [\nu, x \mapsto \true], {\color{blue}E_\text{left}}\right)$\\
 \vspace*{0.1cm}
 
 {\color{blue}\Comment{compute error budget for right DFS-branch $E_\text{left}'$}\vspace*{-1mm}
 \lForEach{$t_i \in \text{targets}(D)$}{$E_{\text{right}}[t_i] \gets \frac{E[t_i]}{2} + E_\text{left}'[t_i]$}}\\
 \vspace*{0.1cm}
 \eIf{$\exists t_i \in \text{targets}(D):~t_i.\text{probupper} - t_i.\text{problower} > {\color{blue}2\varepsilon}$}{
  \Comment{DFS right branch, return the {\color{blue}residual error budget}}
  \Return\FuncSty{dfs}$\left(D, M_\text{right}, [\nu, x \mapsto \false], {\color{blue}E_\text{right}} \right)$
 }{
  \Comment{all probability bounds reached $\varepsilon$-approx., no right DFS}
  \Return {\color{blue}$E_\text{right}$}
 }
}
\caption{Exact and approx. compilation of network\label{alg:network-dfs}}
\end{scriptsize}\end{algorithm}

\begin{figure}
  \begin{center}%
    \ifbool{enabletikz}{\begin{small}\begin{tikzpicture}[scale=0.75,transform shape,->,>=stealth',shorten >=1pt,shorten <= 1pt,node distance=1cm,style={draw,text centered},
main node/.style={draw,thick},
abstract edge/.style={densely dotted},
compilation target/.style={draw=blue!70,thick},
masked true/.style={draw=darkgreen!60,fill=green!20},
masked false/.style={draw=red!60,fill=red!20}
]
  \node[main node,masked true] (x0) {$x_0$};
  \node[main node,masked true] (x1) [right=2cm of x0] {$x_1$};
  \node[main node] (x2) [right=2cm of x1] {$x_2$};
  \node[main node] (x3) [right=2cm of x2] {$x_3$};

  \node[main node,masked true] (o0) [above=0.5cm of x0] {$\phiobj{0}~:~\vee$};
  \node[main node,masked true] (o1) [above=0.5cm of x1] {$\phiobj{1}$};
  \node[main node] (o2) [above=0.5cm of x2] {$\phiobj{2}$};
  \node[main node,masked false] (o3) [above=0.5cm of x3] {$\phiobj{3}~:~\wedge$};
 
  % Variables in object expressions
  \path
    (x0) edge node {} (o0)
    (x1) edge node {} (o1)
    (x1) edge [dashed] node {} (o3)
    (x2) edge node {} (o0)
    (x2) edge node {} (o2)
    (x3) edge node {} (o3);
% 
  % medoid expressions
  \node[main node] (medoid0) [above=0.7cm of o1] {$M^0~:~\Sigma$};
  \node[main node] (medoid1) [above=0.7cm of o2] {$M^1~:~\Sigma$};
% 
%   
  % Object expressions in medoid expressions
  \path[abstract edge]
    (o0) edge node {} (medoid0)
    (o1) edge node {} (medoid0)
    (o2) edge node {} (medoid0)
    (o3) edge node {} (medoid0)
    (o0) edge node {} (medoid1)
    (o1) edge node {} (medoid1)
    (o2) edge node {} (medoid1)
    (o3) edge node {} (medoid1);

  % object assignment
  \node[compilation target] (assign_0_0) [above=1.87cm of o0] {InCl$^{0,0}~:~\wedge$};
  \node[compilation target] (assign_1_0) [right=2cm of assign_0_0] {InCl$^{1,0}~:~\wedge$};
  \node[masked false, compilation target] (assign_1_3) [above=1.87cm of o3] {InCl$^{1,3}~:~\wedge$};
% 
  % medoid expressions used for object assignment
  \path[abstract edge]
    (medoid0) edge node {} (assign_0_0)
    (medoid0) edge node {} (assign_1_0)
    (medoid0) edge node {} (assign_1_3)
    (medoid1) edge node {} (assign_0_0)
    (medoid1) edge node {} (assign_1_0)
    (medoid1) edge node {} (assign_1_3);
    
  \path
    (o3) edge node {} (assign_1_3)
    (o0) edge node {} (assign_0_0)
    (o0) edge[bend left=25] node {} (assign_1_0);
% 
  % Dotted 'edges' to make clear nodes are missing
  \path[-,loosely dotted,thick,shorten >= 3mm, shorten <= 3mm]
    %(medoid1_1_1) edge node {} (medoid1_2_1)
    %(medoid1_2_1) edge node {} (medoid1_k_n)
    (assign_0_0) edge node {} (assign_1_0)
    (assign_1_0) edge node {} (assign_1_3);
    
\end{tikzpicture}\end{small}}{}%
    \caption{Simplified example of an event network.\label{fig:example-dag}}%
  \end{center}\vspace*{-3mm}%
\end{figure}
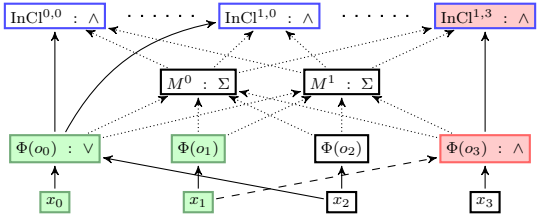

The event programs consist of interconnected events; which are represented
in an \emph{event network}: a graph representation of the
event programs, in which nodes are, e.g., Boolean connectives,
comparisons, aggregates, and c-values. An example of such a network is
depicted in \figref{example-dag}.

The goal is to compute probabilities for the top nodes in the network,
which are referred to as \emph{compilation targets}. These nodes represent events such as ``object $o_i$ is assigned to cluster $C_j$ in iteration $t$''. We keep lower and
upper bounds for the probability of each target.  Initially, these
bounds are $[0,1]$ and they eventually converge during computation.

The bulk-compilation procedure is based on Shannon expansion: select
an input random variable $x$ and partially evaluate each compilation target
$\Phi$ to $\Phi|_{x}$ for $x$ being set to \true ($\top$) and
$\Phi|_{\neg x}$ for $x$ being set to \false ($\bot$). Then, the
probability of $\Phi$ is defined by $\Pr[\Phi] = \Pr[x] \cdot \Pr[\Phi|x] + \Pr[\neg x] \cdot \Pr[\Phi|x]$. We are now left with two simpler events $\Phi|_{x}$
and $\Phi|_{\neg x}$. By repeating this procedure,we eventually resolve all variables in the events to the constants \true or \false.
The trace of this repeated expansion is the decision tree. We need not
materialise the tree. Instead, we just explore it depth-first and
collect the probabilities of all visited branches as well as record
for each event $\Phi$ the sums $L(\Phi)$ and $N(\Phi)$ of
probabilities of those branches that satisfied and respectively did
not satisfy the event. At any time, $L(\Phi)$ and $1-N(\Phi)$
represent a lower bound and respectively an upper bound on the
probability of $\Phi$. This compilation procedure needs time
polynomial in the network size (and in the size of the input data
set), yet in worst case (unavoidably) exponential in the number of
variables used by the events.

For practical reasons, we do not construct $\Phi|_{x}$ and
$\Phi|_{\neg x}$ explicitly, but keep minimal information that, in
addition to the network, can uniquely define them. The process of
computing this minimal information is called {\em masking}. We achieve
this by traversing the network bottom-up and remembering the nodes
that become \true or \false given the values of their children.  When a
compilation target $t$ is eventually masked by a variable assignment
$\nu$, the probability $\Pr(\nu)$ is added to its lower bound if
$\nu(t) = \top$, or subtracted from its upper bound if $\nu(t) =
\bot$.  If one or more targets are left unmasked, a next variable $x'$
is chosen and the process is repeated with $\nu' = \nu\cup \{x'\mapsto
c\}$, where $c$ is either $\top$ or $\bot$.  The algorithm chooses a
next variable $x'$ such that it influences as many events as possible.

Once all compilation targets are masked by an assignment $\nu$, the compilation
backtracks and selects a different assignment for the most recently
chosen variable whose assignments are not exhausted. When all branches
of the decision tree have been investigated, the probability bounds of
the targets have necessarily converged and the algorithm terminates.

\begin{example}\label{ex:dag}
  \figref{example-dag} shows a simplified event network under the
  assignment $\{x_0\mapsto\top,x_1\mapsto\top\}$. The masks of $x_0$
  and $x_1$ are propagated to event nodes $\phiobj{0},~\phiobj{1},~\phiobj{3}$,
  which are now also masked. The red nodes are masked for $\bot$,
  whereas the green nodes are masked $\top$.\punto
\end{example}

\algref{network-dfs} gives the pseudocode for the DFS-traversal of the
decision tree.  The blue lines are necessary for approximate probability
computation and will be explained later. Compilation starts with an
empty branch (variable assignment) $\nu$; the mask values $M[v_i]$ and
probability bounds for all nodes in the event network are initialised.
The error budgets $E[t_i]$ for the targets are set to $0$ for exact
computation. After the initialisation, the \textsc{dfs} procedure is
called using $\nu = \{\,\}$. The procedure selects the first
variable $x$ and recursively call itself using two newly created
branches of the decision tree: one for $\nu = \{x \mapsto \top\}$ and
one with $\nu = \{x \mapsto \bot\}$. These branches are propagated
into the event network using the \textsc{Mask} procedure. If every
target is reached, \textsc{dfs} returns. Otherwise, it selects a
next variable $x$ and recursively calls \textsc{dfs} on the two new tree branches.

\algref{network-mask} performs mask propagation: a mask (assignment) for a variable
$x$ is inserted into the network, and the variable node propagates the
mask to its parent nodes. Depending on the event node, its node mask is either
updated and propagated further, or propagation is stopped in case a
mask cannot be established.

Convergence of the algorithm (\eg, clustering) can be detected by comparing the mask values at network
nodes corresponding to iteration $t$ with the masks of nodes for
iteration $t+1$. If none of the mask assignments has changed between iterations, then the algorithm has converged.

\subsection{Bounded-range loops in event networks}

Event programs can contain bounded-range loops for iterative
algorithms. ENFrame offers two ways of encoding such loops in an event
network: \emph{unfolded}, in which case the events at any loop
iteration are explicitly stored as distinct nodes in the network, or a
more efficient \emph{folded} approach in which all iterations are
captured into a single set of nodes. The compilation of the network
then involves looping. The pseudocode in Algorithms~\ref{alg:network-dfs} and \ref{alg:network-mask} 
assumes unfolded event networks. They need minor modifications to work
on folded networks: the mask data structure $M$ becomes
two-dimensional to be able to store the mask for a node $v$ at any
iteration $t$ ($M[t][v]$) the \textsc{dfs} procedure needs an
additional parameter $t$ for the current compilation iteration, and the network
requires an additional node to perform the transition from iteration
$t$ to iteration $t+1$.  The extra logic required for the
\textsc{mask} function is:

\begingroup\ignorelatexerrors % required because we use [H] with \begin{algorithm} whilst in two-column mode...
\noindent\hspace*{0.05\linewidth}\begin{minipage}{0.95\linewidth}\vspace*{1mm}
\RestyleAlgo{plain}\begin{algorithm}[H]\begin{scriptsize}\SetInd{1.6mm}{1.6mm}
\DontPrintSemicolon
\Case(\Comment*[f]{loop node}){$\triangledown$}{
 $M[t+1][v] \gets M[t][c]$\Comment*{carry over mask to next iteration}
 $t \gets t + 1$\Comment*{increase iteration counter}
}
\end{scriptsize}\end{algorithm}\end{minipage}
\endgroup

Additionally, probability bounds of compilation targets should only be
updated if $t$ is the last iteration, and propagation should only take
place if $t$ is \emph{not} the last iteration.

\subsection{Approximation with error guarantees}
The compilation procedure can be extended to achieve an anytime absolute \eapprox with error guarantees. The idea is to stop the probability computation as soon as the bounds of all compilation targets are sufficiently tight.

\begin{definition}
Given a fixed error $0\leq\varepsilon\leq 1$ and events
$(t_0,\ldots,t_{n-1})$ with probabilities $(p_0,\ldots,p_{n-1})$. An absolute
\eapprox for these events is defined as a tuple
$(\hat{p}_0,\ldots,\hat{p}_{n-1})$ such that $\forall 0\leq i < n:
p_i-\varepsilon\leq \hat{p}_i\leq p_i+\varepsilon.$\punto
\end{definition}

The compilation of the network yields probability bounds
$([L_0,U_0],\ldots,[L_{n-1},U_{n-1}])$ for the targets $(t_0,\ldots,t_{n-1})$. It
can be easily seen that an absolute \eapprox can be defined by any tuple $(\hat{p}_0,\ldots,\hat{p}_{n-1})$ such that $\forall
0\leq i < n: U_i-\varepsilon\leq \hat{p}_i\leq L_i+\varepsilon.$ We
thus need to run the algorithm until $U_i-L_i\leq 2\varepsilon$ for
each target $t_i$.

There exist multiple strategies for investing this $2\varepsilon$ error budget for every target. We next discuss three such strategies. The \texttt{lazy}
scheme follows the exact probability computation approach and stops as
soon as the bounds become tight enough.  Effectively, this results in
investing the entire error budget into the rightmost branches of the
decision tree.  The \texttt{eager} scheme spends the entire error
budget as soon as possible, i.e., on the leftmost branches of the
decision tree, and then continues as in the case of exact computation.
At each node in the decision tree, the \texttt{hybrid} scheme divides
the current error budget equally over the two branches. Any residual,
unused error budget is transferred to the next branch.

The blue lines in \algref{network-dfs} show how the \textsc{dfs}
procedure can be extended to support anytime absolute \eapprox with
error guarantees using the \texttt{hybrid} scheme. The \textsc{dfs}
procedure is called using a non-zero error budget $2\varepsilon$, and
it assigns half of the budget to the newly created left branch of the
decision tree. The recursive \textsc{dfs} call returns the residual
error budget of each target, which is then added to the budget for the
right branch.

\subsection{Distributed probability computation}

By splitting the task of exploring the decision tree in a number of
jobs, the compilation can be performed concurrently by multiple
threads or machines.  A worker explores a tree fragment of a given maximum size. For simplicity, we define the size of a job to be the
depth $d$ of the sub-tree to explore. The computation then proceeds as
follows. One worker explores the tree from the root and every time it
reaches depth $d$, it forks a new job that continues from that node as
its root.  \nop{At creation, the new job gets the mask at its root
  node from the parent job.} Given that the maximum depth of the tree
is the number of variables $m$, the number of jobs created is at most
$\overset{\lceil\frac{m}{d}\rceil-1}{\underset{i=0}{\Sigma}} 2^{i\cdot
  d}$, where the cost of each job would propagate at most $2^d$ variable valuations $\nu$ into the event network. Each job incurs the cost of communicating the mask at job
creation and the probability bounds for each target at the end of the
job. In case of approximation, the error budgets need to be synchronised both at the start and end of a job.

\begin{algorithm}[t]\begin{scriptsize}\SetInd{1.6mm}{1.6mm}
%todo:
% - make 'targets reached' and 'targets approximated' more formal
% - 
\DontPrintSemicolon
\FuncSty{Mask}(\DataSty{network $D$, masks $M$, node $v$, child $c$, prob $p$})
\Begin{

\Switch{$v.\text{nodetype}$}{
 \lCase(\Comment*[f]{$M[c] \in \mathbb{B},~M[v] \in \mathbb{B}$}){$\neg~~:$}{$M[v] \gets \neg M[c]$}
 \Case(\Comment*[f]{$M[c] \in \mathbb{B},~M[v] \in \mathbb{B}$}){$\wedge$}{
  \lIf{$M[c] = \algfalse$}{$M[v] \gets \false$}\\
  \lElseIf{$\forall c_i \in v.\text{children}: M[c_i] = \algtrue$}{$M[v] \gets \true$}
 }
 \Case(\Comment*[f]{$M[c] \in \mathbb{B},~M[v] \in \mathbb{B}$}){$\vee$}{
  \lIf{$M[c] = \algtrue$}{$M[v] \gets \true$}\\
  \lElseIf{$\forall c_i \in v.\text{children}: M[c_i] = \algfalse$}{$M[v] \gets \false$}
 }
 \Case(\Comment*[f]{c-value: $M[c] \in \mathbb{B},~M[v] \in \mathbb{R}$}){$\otimes$}{
  \Comment{update lower or upper bound of c-value ($\mathbb{R}$)}
  \lIf{$M[c] = \algtrue$}{$M[v].\text{\emph{lower}} \gets v.\text{\emph{val}}$}\\
  $M[v].\text{\emph{upper}} \gets M[v].\text{\emph{lower}}$
 }
 \Case(\Comment*[f]{sum of c-values: $M[c] \in \mathbb{R},~M[v] \in \mathbb{R}$}){$\Sigma$}{
  \Comment{update lower or upper bound of c-values ($\mathbb{R}$)}
  $M[v].lower \gets M[v].lower + M[c].lower$\\
  $M[v].upper \gets M[v].upper - (c.val - M[c].lower)$\\
 }
 \Case(\Comment*[f]{$M[c] \in \mathbb{R},~M[v] \in \mathbb{B}$}){$<$}{
  % left_upper <= r_lower  ==> TRUE
  % left_lower > r_upper ==> FALSE
  \lIf{$v.\text{left}.\text{upper} < v.\text{right}.\text{lower}$}{$M[v] \gets \true$}\\
  \lElse{$v.\text{left}.\text{lower} \geq v.\text{right}.\text{upper}$}{
   $M[v] \gets \false$
  }
 }
}
\vspace*{0.1cm}

\If{$\left(M[v] \in \mathbb{B} \text{\textbf{\emph{ and }}} M[v] \neq \text{unknown}\right)$ \textbf{or} $M[v] \in \mathbb{R}$}{
 \If{$v \in \text{targets}(D)$}{
  \Comment{$v$ is compilation target: update probability bounds}
  \lIf{$M[v] = \true$}{$v.\text{\emph{problower}} \gets v.\text{\emph{problower}} + p$}\\
  \lElse{$v.\text{\emph{probupper}} \gets v.\text{\emph{probupper}} - p$}
 }
 \Comment{propagate mask to parents of $v$}
 \ForEach{$\text{p}_i \in v.\text{parents}$}{
  \Comment{check whether $p_i$ is already fully masked}
  \If{$\left(M[p_i] \in \mathbb{B}~\text{\textbf{and}}~M[p_i] = \text{unknown}\right)$ \textbf{or} \\
    $~\,\;\left(M[p_i] \in \mathbb{R}~\text{\textbf{and}}~M[p_i].\text{upper} \neq M[p_i].\text{lower}\right)$}{$M \gets $\FuncSty{mask}($D,M,p_i,v,p$)} 
 }
}
\vspace*{0.1cm}
\Return{$M$}
}
\caption{Masking of nodes in an event network\label{alg:network-mask}}
\end{scriptsize}\end{algorithm}

\ifbool{renderplots}{%%% First: some PGF settings:

\ifbool{renderplots}{% plot rendering is enabled
\usetikzlibrary{spy}
\pgfplotstableset{col sep=semicolon}

% Colors for experiment 3a: 3 algorithm types, with 2 types of data each
\pgfplotscreateplotcyclelist{icml-exp3a-3x2}{%
  {red,mark=10-pointed star},
  {blue,mark=diamond*},
  {darkgreen,mark=square*,mark options={scale=0.7}},
  {densely dashed,red,mark=10-pointed star},
  {densely dashed,blue,mark=diamond*},
  {densely dashed,darkgreen,mark=square*,mark options={scale=0.7}}%
}

% Colors for experiment 3a: 4 algorithm types, with 2 types of data each
\pgfplotscreateplotcyclelist{icml-exp3a-4x2}{%
  {red,mark=10-pointed star},
  {blue,mark=diamond*},
  {darkgreen,mark=square*,mark options={scale=0.7}},
  {lightgreen,mark=triangle*},
  {densely dashed,red,mark=10-pointed star},
  {densely dashed,blue,mark=diamond*},
  {densely dashed,darkgreen,mark=square*,mark options={scale=0.7}},
  {densely dashed,lightgreen,mark=triangle*}%
}

% Colors for experiment 3b: 4 algorithm types, with 2 types of data each
\pgfplotscreateplotcyclelist{icml-exp3b-4x2}{%
  {violet,mark=*},
  {orange,mark=diamond*,every mark/.append style={rotate=90}},
  {darkgreen,mark=square*,mark options={scale=0.7}},
  {lightgreen,mark=triangle*},
  {densely dashed,red,mark=halfcircle*,mark options={solid},every mark/.append style={rotate=180}},
  {densely dashed,blue,mark=halfdiamond*,mark options={solid}},
  {densely dashed,darkgreen,mark=halfsquare*,mark options={solid}},
  {densely dashed,lightgreen,mark=triangle*,mark options={solid},every mark/.append style={rotate=180}}
}

% Colors for experiment 3b (ds1): 4 algorithm types, with 3 types of data each
\pgfplotscreateplotcyclelist{icml-exp3b-4x3}{%
  {violet,mark=*},
  {orange,mark=diamond*,every mark/.append style={rotate=90}},
  {darkgreen,mark=square*,mark options={scale=0.7}},
  {lightgreen,mark=triangle*},
  {densely dashed,violet,mark=halfcircle*,mark options={solid},every mark/.append style={rotate=180}},
  {densely dashed,orange,mark=halfdiamond*,mark options={solid},every mark/.append style={rotate=90}},
  {densely dashed,darkgreen,mark=halfsquare*,mark options={solid}},
  {densely dashed,lightgreen,mark=triangle,mark options={solid}},
  {densely dotted,violet,mark=o,mark options={solid}},
  {densely dotted,orange,mark=diamond,mark options={solid},every mark/.append style={rotate=90}},
  {densely dotted,darkgreen,mark=square,mark options={solid}},
  {densely dotted,lightgreen,mark=triangle,mark options={solid},every mark/.append style={rotate=180}}%
}

% Colors for experiment 3b (ds1): 3 algorithm types, with 3 types of data each
\pgfplotscreateplotcyclelist{icml-exp3b-3x3}{%
  {violet,mark=*},
  {orange,mark=diamond*,every mark/.append style={rotate=90}},
  {darkgreen,mark=square*,mark options={scale=0.7}},
  {densely dashed,violet,mark=halfcircle*,mark options={solid},every mark/.append style={rotate=180}},
  {densely dashed,orange,mark=halfdiamond*,mark options={solid},every mark/.append style={rotate=90}},
  {densely dashed,darkgreen,mark=halfsquare*,mark options={solid},mark size=2.2},
  {densely dotted,violet,mark=o,mark options={solid}},
  {densely dotted,orange,mark=diamond,mark options={solid},every mark/.append style={rotate=90}},
  {densely dotted,darkgreen,mark=square,mark options={solid,scale=0.7}}%
}

\pgfplotscreateplotcyclelist{naive-exact-hybrid-lazy-eager-hybridd-numvars}{%
  {red,mark=10-pointed star}, % naive
  {blue,mark=diamond*}, % exact
  {darkgreen,mark=square*,mark options={scale=0.7}}, % hybrid
  {violet,mark=*}, % lazy
  {orange,mark=diamond*,every mark/.append style={rotate=90}}, % eager
  {lightgreen,mark=triangle*}, % hybrid-d
  {gray,dashed,mark size=1,mark=*}% % num-variables
}

\newcommand{\pgfplotsfontstyle}{\tiny}

\pgfplotsset{every axis label={font=\pgfplotsfontstyle},
  tick label style={font=\pgfplotsfontstyle},
  label style={font=\pgfplotsfontstyle},
  legend style={font=\pgfplotsfontstyle,row sep=-0.1cm,cells={anchor=west}},
  title style={font=\pgfplotsfontstyle,align=center},
  scale only axis,
  table/header = false,
  xtick pos=left,
  every axis/.append style={mark size=1.75}
}

% #1 = y-value for timeout bar (e.g. 3600)
% #2 = location of label (e.g. 'west' or 'east')
\newcommand{\pgfaddtimeoutline}[2]{
  \draw[red,densely dotted] (axis cs:\pgfkeysvalueof{/pgfplots/xmin},#1) -- (axis cs:\pgfkeysvalueof{/pgfplots/xmax},#1);
  \ifthenelse{\equal{#2}{west}}%
%   {\node at (axis cs:\pgfkeysvalueof{/pgfplots/xmin},#1) [color=red,fill=white,anchor=west,font=\tiny] {timeout = #1 sec.};}
%   {\node at (axis cs:\pgfkeysvalueof{/pgfplots/xmax},#1) [color=red,fill=white,anchor=east,font=\tiny] {timeout = #1 sec.};}
   {\draw (axis cs:\pgfkeysvalueof{/pgfplots/xmin},#1) ++ (3pt, -1pt) node [color=red,fill=white,anchor=west,scale=0.5] {\parbox{3cm}{\vspace*{-5pt}timeout = #1 sec.}};}
   {\draw (axis cs:\pgfkeysvalueof{/pgfplots/xmax},#1) ++ (-3pt, -1pt) node [color=red,fill=white,anchor=east,font=\tiny] {\parbox{3cm}{\vspace*{-5pt}timeout = #1 sec.}};}
}

% #1 = only show data points with these indices (optional, comma-separated list. Example: 5,6,7)
% #2 = additional options for \addplot (optional, e.g. 'forget plot')
% #3 = data filename
% #4 = column index for x value
% #5 = column index for y value
% #6 = column index for lower bound of y value
% #7 = column index for upper bound of y value
%\newcommand{\adderrorbarsplot}[5]{
%\newcommand{\adderrorbarsplot}[6][]{
\DeclareDocumentCommand{\adderrorbarsplot}{O{} O{} m m m m m}{
  \ifstrempty{#1}{%
    \addplot+[solid,#2,forget plot,only marks,error bars/.cd,y dir=minus,y explicit] table [x index=#4, y expr=\thisrowno{#5}/1000, y error expr=\thisrowno{#5}/1000-\thisrowno{#6}/1000]{#3};
    \addplot+[solid,#2,forget plot,only marks,error bars/.cd,y dir=plus,y explicit] table [x index=#4, y expr=\thisrowno{#5}/1000, y error expr=\thisrowno{#7}/1000-\thisrowno{#5}/1000]{#3};
    \addplot+[#2] table [x index=#4, y expr=\thisrowno{#5}/1000]{#3};
  }{ % apply filter
    \addplot+[filter x indices={#1},mark indices={200},solid,forget plot,only marks,error bars/.cd,y dir=minus,y explicit] table [x index=#4, y expr=\thisrowno{#5}/1000, y error expr=\thisrowno{#5}/1000-\thisrowno{#6}/1000]{#3};
    \addplot+[filter x indices={#1},mark indices={200},solid,forget plot,only marks,error bars/.cd,y dir=plus,y explicit] table [x index=#4, y expr=\thisrowno{#5}/1000, y error expr=\thisrowno{#7}/1000-\thisrowno{#5}/1000]{#3};
    \addplot+[mark indices={#1}] table [x index=#4, y expr=\thisrowno{#5}/1000]{#3};
  }
}

\newcommand{\filterloopdo}[2]{
  %\typeout{got item: #1. Does it equal #2 from the list?}
  \def\xcoordinate{#1}
  \def\coordinatelistitem{#2}
  \ifnumcomp{\xcoordinate}{=}{\coordinatelistitem}{\typeout{including x=\xcoordinate}\xdef\filterloopdoresult{\xcoordinate}}{}
}

%  #1 : the x-coordinate of the data point
%  #2 : csv list of markers to shrink
\newcommand{\shrinkmarkersxindexmarkersize}[2]{4}

% Can be used in conjunction with \addplot, for example:
%   \addplot [filter x coordinates={25,30,200,500}] table [x index=1, y expr=\thisrowno{24}/1000]{\tableC};
%
% Which will filter the data used to the specified x-coordinates.
%
% If this makes the legend look ugly, use the following inside the \begin{tikzpicture}...\end{tikzpicture} block:
%   \pgfplotsset{legend image code/.code={\draw[mark indices={},mark repeat=2,mark phase=2] plot coordinates {(0cm,0cm)(0.3cm,0cm)(0.6cm,0cm)};}}
\pgfkeys{
    /pgfplots/filter x coordinates/.style={
	/pgfplots/x filter/.code={%
		\def\inputxcoordinate{\pgfmathresult}%
		\xdef\filterloopdoresult{-1}
		\csvloop[\filterloopdo{\inputxcoordinate}]{#1}
		\ifnum\filterloopdoresult<0
		  \def\pgfmathresult{}
		\fi
	}
    },/pgfplots/filter x coordinates/.default={},
    /pgfplots/filter x indices/.style={
	/pgfplots/x filter/.code={%
		\FPeval{\inputxindex}{clip(\coordindex+1)}
		\xdef\filterloopdoresult{-1}
		\csvloop[\filterloopdo{\inputxindex}]{#1}
		\ifnum\filterloopdoresult<0
		  \def\pgfmathresult{}
		\fi
	}
    },/pgfplots/filter x indices/.default={}
}
}{}% end: \ifbool{renderplots}
}{}

\setlength{\intextsep}{0cm}
\setlength{\floatsep}{0cm}
\setlength{\textfloatsep}{2mm}
\newlength{\defaultplotwidth}\setlength{\defaultplotwidth}{7.35cm}

\section{Experimental evaluation}\label{sec:experiments}

%%%%%%%%%%%%%%%%%%%%%%%%%%%%%%%%%%%%%%%%
\begin{figure*}[t]
\noindent\begin{minipage}{0.5\linewidth}\includeplot[width=0.6\textwidth,height=2.5cm]{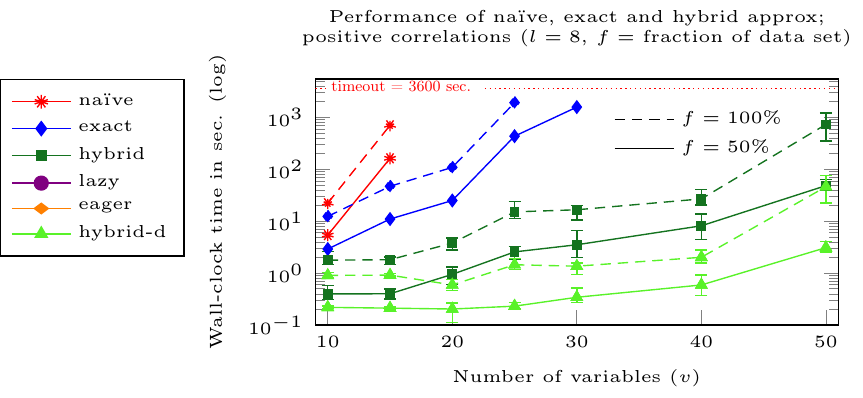}\end{minipage}%
\hfill\begin{minipage}{0.5\linewidth}\includeplot[width=7.55cm,height=2.5cm]{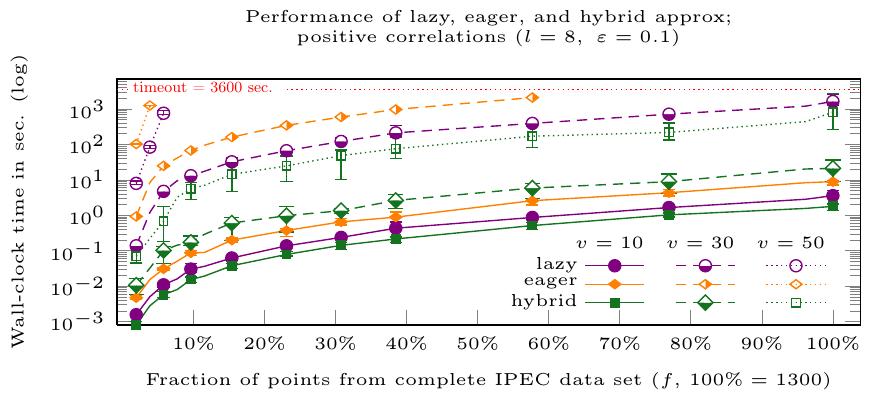}\end{minipage}
\vspace*{-4mm}\caption{Positively correlated data. On the left: scalability in terms of variables, on the right: scalability of approximations in terms of size of the data set (\texttt{hybrid-d} not shown for visibility).\label{fig:exp3-ds1}}%
\vspace*{-2mm}%
\end{figure*}
%%%%%%%%%%%%%%%%%%%%%%%%%%%%%%%%%%%%%%%%

%%%%%%%%%%%%%%%%%%%%%%%%%%%%%%%%%%%%%%%% FIGURE MUTEX AND CONDITIONAL DATA
\begin{figure*}[t]%
\noindent\begin{minipage}{8.5cm}\includeplot[width=6.5cm,height=2.5cm]{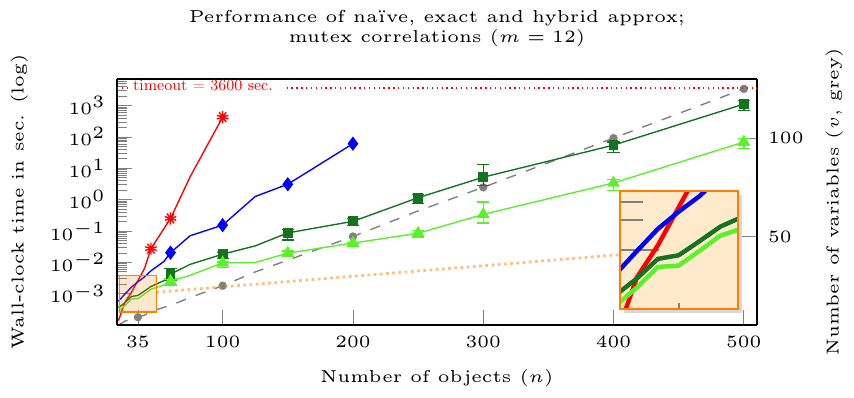}\end{minipage}%
\hfill\begin{minipage}{8.5cm}\includeplot[width=6.5cm,height=2.5cm]{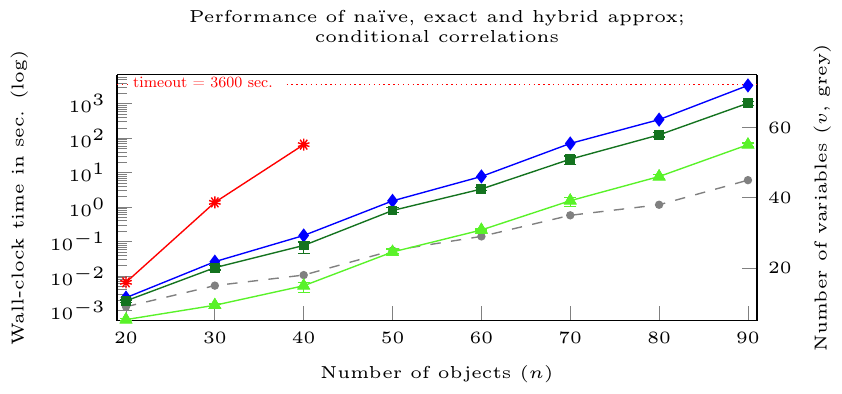}\end{minipage}%
\vspace*{-3mm}
\caption{Mutex and conditionally correlated data (legend: see Fig. \ref{fig:exp3-ds1}). Algorithms {\tt eager} and {\tt lazy} overlap with {\tt exact}, and are not shown. Grey dashed line indicates number of variables.}
\label{fig:exp3-ds23}\vspace*{-4mm}
\end{figure*}
%%%%%%%%%%%%%%%%%%%%%%%%%%%%%%%%%%%%%%%%

This section describes an experimental evaluation of clustering probabilistic data using ENFrame. The focus of this evaluation is a preliminary benchmark of the performance of the probability computation algorithms introduced in \secref{probcomp}. At the end of this section, we comment on further experimental considerations that could not be included in full due to space limitations.

%%%%%%%%%%%%%%%%%%%%%%%%%%%%%%%%%%%%%%%%
{\noindent\bf Data.} We use a data set describing network load and occurrences of \emph{partial discharge} in energy distribution networks~\cite{Michel:2011}. This data is gathered from two different types of sensors: partial discharge sensors installed on switchgear and cables in substations of the distribution network, and network load sensors in substations. We aggregate the number of partial discharge occurrences over the duration of an hour and subsequently pair this value with the average network load during that hour. Clustering can assist in detecting anomalies and predicting failures in the energy networks.

%%%%%%%%%%%%%%%%%%%%%%%%%%%%%%%%%%%%%%%%
{\noindent\bf Uncertainty.} Our goal is to show that ENFrame can deal with common correlations patterns that occur in probabilistic data \cite{Agrawal:2006:trio,SenDeshpande:2007:correlations,SORK:2011}. Each data point is associated with an event described by Boolean random variables, whose probabilities for true are chosen at random from the range $[0.5, 0.8]$. Different values would make the probabilities of clustering events too close to 0 or 1 which are then easily approximable. The experiments were carried out using three types of correlations to illustrate ENFrame's capability to process arbitrarily correlated data.

The \emph{positive correlations} scheme yields events such that two data points are either positively correlated or independent. Each event is a disjunction of $l$ distinct positive literals.
In the {\em mutex correlations} scheme, the data points are partitioned in mutex sets of cardinality (at most) $m$: any two points are mutually exclusive within a mutex set and independent across the sets. The \emph{conditional correlations} scheme expresses uncertainty as a Markov chain, using one node per data point. Let $\Phi_i$ be the event that the data point $o_i$ exists.  The event $\Phi_{i+1}$ becomes $(\Phi_i\wedge x_{i+1}^{\text{t}}) \vee (\neg\Phi_i\wedge x_{i+1}^{\text{f}})$; it is a disjunction of two events, for the cases that $o_i$ exists or not. We thus introduce two new Boolean random variables $x_{i+1}^{\text{t}}$ and $x_{i+1}^{\text{f}}$ per data point $o_{i+1}$. For every correlation scheme, a \emph{group size} of 4 has been used, \ie data points were divided in groups with identical lineage. This is realistic for uncertain time-series sensor data: readings from a small time window have identical correlations and uncertainty. Additionally, we show experiments with a varying fraction of \emph{certain} data points.

%%%%%%%%%%%%%%%%%%%%%%%%%%%%%%%%%%%%%%%%
{\noindent\bf Algorithms.} We report on performance benchmarks for $k$-medoids clustering on the energy network data set, comparing ENFrame to \texttt{na\"ive} clustering. The \texttt{na\"ive} approach computes an equivalent clustering by explicitly iterating over all possible worlds. We show the performance of multiple probability computation algorithms of ENFrame: the sequential \texttt{exact} approach, three sequential approximation schemes (\texttt{eager}, \texttt{lazy}, \texttt{hybrid}), and distributed hybrid approximation (\texttt{hybrid-d}). All approximation algorithms are set to compute probabilities with an (absolute) error of at most $\varepsilon = 0.1$, the compilation targets are the events that represent medoid selection.

Algorithms described in the literature (see Section~\ref{sec:related}) simplify the clustering problem by ignoring correlations, using expected distances, and producing a deterministic output. They might outperform our sequential algorithms, at the cost of producing an output that can be arbitrarily off from the golden standard of clustering in each world. Unfortunately, none of the reported prototypes was available for testing at the time of writing.

%%%%%%%%%%%%%%%%%%%%%%%%%%%%%%%%%%%%%%%%
{\noindent\bf Setup.} The experiments were carried out on an Intel
Xeon X5660/ 2.80GHz machine with 4GB of RAM, running Ubuntu with Linux kernel 3.5. The timings reported for \texttt{hybrid-d} were obtained by simulating distributed computation on a single machine. The algorithms are implemented in C++ (GCC 4.7.2). Each plot in Figures~\ref{fig:exp3-ds1} and \ref{fig:exp3-ds23} depicts average performance with min/max ranges of five runs with randomly generated event expressions, different probabilities, and three clustering iterations (using Euclidean distance).

%%%%%%%%%%%%%%%%%%%%%%%%%%%%%%%%%%%%%%%%

\parhead*{Sequential algorithms.}
Figures~\ref{fig:exp3-ds1} and \ref{fig:exp3-ds23} show that all of ENFrame's probability computation algorithms outperform the \texttt{na\"ive} algorithm by up to six orders of magnitude for each data set with more than 10 variables. Furthermore, the {\tt hybrid} approximation can be up to four orders of magnitude faster than {\tt exact} computation.

Indeed, for a very small number of possible worlds (i.e., a small number of variables), it pays off to cluster individually in each world and avoid the overhead of the event networks. For a larger number of worlds, our exact and approximate approaches quickly become up to six orders of magnitude faster.  The \texttt{na\"ive} method times out for over 25 variables in every correlation scheme.

The reason why our approximation schemes outperform \texttt{exact} is
as follows. For a given depth $d$, there are up to $2^d$ nodes in the
decision tree that contribute to the probability mass of a node in the
event network. The contributed mass decreases exponentially with an
increase in depth, which suggests that most nodes in the decision tree
only contribute a small fraction of the overall probability
mass. Depending on the desired error bound, a shallow exploration of
the decision tree could be enough to achieve a sufficiently large
probability mass.

Among the approximation algorithms, {\tt hybrid} performs best; it outperforms \texttt{exact} by up to four orders of magnitude since it does only need to traverse a shallow prefix of the decision tree. The algorithm invests the error budget over the entire width of the decision tree, cutting branches of the tree after a certain depth. The other two methods ({\tt eager} and {\tt lazy}) use the budget to respectively cut the first and last branches, while exploring other branches in full depth.

For positive correlations, {\tt lazy} performs very well, because the
decision tree is very unbalanced under this scheme. The left branches
of the tree correspond to variables being set to \true, which quickly
satisfy the (disjunctive) input events and allow for compilation
targets to be reached. Further to the right, branches correspond to
variables being set to \false. More variables need to be set to
(un)satisfy the disjunctive input event, thus leading to longer
branches. The \texttt{lazy} algorithm saves the error budget until the
very last moment and can therefore prune the deep branches whilst
maintaining the $\varepsilon$-approximation. The decision trees for
the mutex and conditional correlation schemes are more balanced,
resulting in both \texttt{lazy} and \texttt{eager} to perform almost
identically to \texttt{exact}. Hence, they are not shown
in \figref{exp3-ds23}.

%%%%%%%%%%%%%%%%%%%%%%%%%%%%%%%%%%%%%%%%
{\noindent\bf Distributed algorithms.} By distributing the probability computation task, we can significantly increase ENFrame's performance.  Figures~\ref{fig:exp3-ds1} and \ref{fig:exp3-ds23} show the timings for {\tt hybrid-d} using $w=16$ workers and job size $d = 3$ (as detailed below). For all correlation schemes, {\tt hybrid-d} gets increasingly faster than {\tt hybrid} as we increase the number of variables. For small numbers of variables, the overhead of distribution does not pay off. The benefits are best seen for mutex correlations and over 100 objects (over 60 variables), where {\tt hybrid-d} becomes more than one order of magnitude faster than {\tt hybrid}. For readability reasons, the performance of \texttt{hybrid-d} is not depicted in \figref{exp3-ds1} (right); its performance is up to one order of magnitude better than \texttt{hybrid}, as can be seen in \figref{exp3-ds1}. For ten variables, there is only a small performance gain when compared to the single-threaded \texttt{hybrid} approximation: the decision tree remains small, as is the number of jobs that can be generated. However, for $30$ and $50$ variables, \texttt{hybrid-d} yields a performance improvement of more than one order of magnitude when compared to \texttt{hybrid}.

\begin{figure}[t]
\noindent\hfill\includeplot[width=\defaultplotwidth,height=2cm]{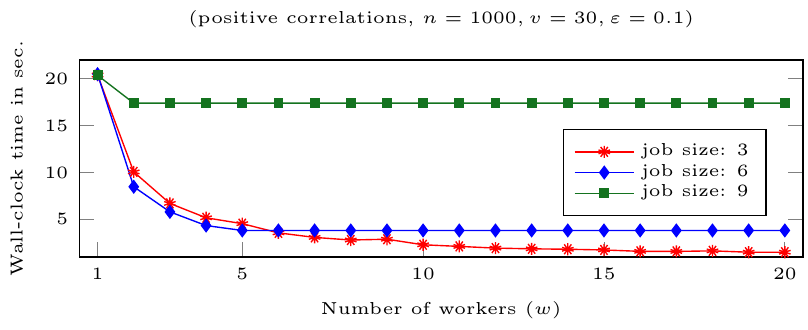}\vspace*{-4.3mm}%
\caption{Performance of distributed probability computation as function of number of workers.\break~\label{fig:distributed}}\vspace*{-3mm}%
\end{figure}

Figure~\ref{fig:distributed} shows the influence of the number of workers on {\tt hybrid-d}'s performance for varying job sizes. A job is the work unit allocated to a worker at any one time; a size of $d$ means that the worker has to explore a fragment of the decision tree of depth at most $d$ and would need to traverse the event network at most $2^d$ times. For large job sizes, the overall number of jobs decreases; in the case of positive correlations, the number of jobs of size 9 is small since the decision tree is very unbalanced and only a few branches on the right-hand side of the tree grow deeper than nine variables. Therefore, increasing the number of workers would not help; indeed, there is no improvement for more than four workers for job sizes larger than 5. However, for a job size of 3, up to 16 workers can still be beneficial. In our experiment, smaller job sizes led to a performance gain of up to one order of magnitude, since they allow for a more equal distribution of the work over the available workers. Synchronisation did not play a significant role in our setup.

%%%%%%%%%%%%%%%%%%%%%%%%%%%%%%%%%%%%%%%%

%%%%%%%%%%%%%%%%%%% FIGURE GENERATED LARGE DATASET WITH CERTAIN DATA

\begin{figure}[t]
\noindent\includeplot[width=\defaultplotwidth,height=2cm]{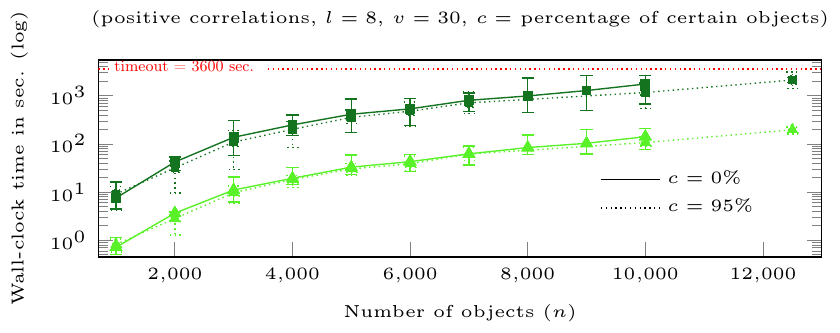}\vspace*{-4mm}%
\caption{Performance of \texttt{hybrid} and \texttt{hybrid-d} on large-scale generated data sets with different fractions of certain data points.\label{fig:large-certain}}%
\end{figure}

%%%%%%%%%%%%%%%%%%%%%%%%%%%%%%%%%%%%%%%%

\parhead*{Certain data points.} \figref{large-certain} shows that the performance improves as the number of certain data points (\ie, objects that occur in all possible worlds) increases. The speedup in such cases is explained by the fact that the distance sums of possible medoids to data points in a cluster become less complex and can be initialised using the distances to objects that certainly exist. Consequently, fewer variables assignments are needed to decide on a cluster medoid, resulting in a shallower decision tree and a speedup in the compilation time.

%%%%%%%%%%%%%%%%%%%%%%%%%%%%%%%%%%%%%%%%
{\noindent\bf Further findings.}  We have investigated the influence of the number of dimensions, data point coordinates, the error budget, the numbers of iterations, and alternative clustering compilation targets on the performance of ENFrame, as well as its total memory usage. As is the case with traditional \kmedoids on certain data, the number of dimensions has no influence on the computation time. The reported performance gap between \texttt{exact} and \texttt{hybrid} shows that performance is highly sensitive to the error budget. The number of iterations has a linear effect on the running time of the algorithm. The number of targets (including targets representing co-occurrence queries) has a minor influence on performance; due to the combinatorial nature of \kmedoids, clustering events are mostly satisfied in bulk and it is thus very rare that one event alone is satisfied at any one time. This also explains why experiments with other types of compilation targets (\eg, object-cluster assignment, pairwise object-cluster assignment) show very similar performance. In our experiments, the size of the event networks grows linearly in the number of objects and clusters and the memory usage of ENFrame is under 1GB.

%%%%%%%%%%%%%%%%%%%%%%%%%%%%%%%%%%%%%%%%
{\noindent\bf Comments on clustering quality.} The research effort
described in this paper is mainly concerned with expressing a rich
class of algorithms for data analysis in ENFrame, and scalability of
the probability computation task; \kmedoids clustering is merely a
use-case to show ENFrame's flexibility and scalability. The adaptation
of \kmedoids to ENFrame has the exact same quality as the ``golden
standard'': $k$-medoids applied in each possible world, yet without
actually explicitly iterating over all possible worlds. This is not
the case for prior work that does not support correlated uncertain
input and uncertain output. An extensive quality comparison is out of
scope of this paper, but is part of future work. A common approach to
assess the quality measure of a probabilistic method like ours, is to
assume a notion of ground truth. It is
unclear how to deal with probabilistic data which represents
inherently contradictory information for which no ground truth exists
or is known.

\section{Related Work}
\label{sec:related}

Our work is at the confluence of several active research areas:
probabilistic data management, data analytics platforms, and
provenance data management.

%%%%%%%%%%%%%%%%%%%%%%%%%%%%%%%%%%%%%%%%%%% FIG FROM EXPERIMENTS
%%%%%%%%%%%%%%%%%%%%%%%%%%%%%%%%%%%%%%%%%%%

%%%%%%%%%%%%%%%%%%%%%%%%%%%%%%%%%%%%%%%%%%%
{\noindent\bf Probabilistic data mining and querying.} Our work adds
to a wealth of literature on this topic~\cite{Aggarwal:2009,SORK:2011}
along two directions: distributed probability computation techniques
and a unified formalisation of several clustering algorithms in line
with work on probabilistic databases.

Distributed probability computation has been approached so far only in
the context of the SimSQL/MCDB system, where approximate query results are
computed by Monte Carlo simulations~\cite{MCDB:TODS:2011,Zhuhua:2013:SimSql}. This contrasts with our approach in
that MCDB was not designed for exact and approximate computation with
error guarantees and does not exploit at runtime
symbolically-represented correlations allowed by pc-tables and by
ENFrame.

Early approaches to mining uncertain data are based on imprecise
(fuzzy) data, for example using intervals, and produce fuzzy (soft)
and hard output. Follow-up work shifted to representation of
uncertainty by (independent) probability density functions per data
point. In contrast, we allow for arbitrarily correlated discrete
probability distributions. The importance of correlations has been previously acknowledged for clustering~\cite{Volk:2009:ClusteringWorlds} and frequent pattern mining~\cite{Sun:2010:frequentpattern}.
A further key aspect of our approach that
is not shared by existing uncertain data mining approaches is that we
follow the possible worlds semantics throughout the whole mining
process. This allows for exact and approximate computation with error
guarantees and sound semantics of the mining process that is
compatible with probabilistic databases.  This cannot be achieved by
existing work; for instance, most existing $k$-means clustering
approaches for uncertain data define cluster centroids using {\em
expected distances} between data
points~\cite{Chau06uncertaindata,Ngai:2006,Gullo:2008,Kriegel:2005,Gullo:2008b,kao:2010}
or the {\em expected variance} of all data points in the same
cluster~\cite{Gullo:2010}; they also compute hard clustering where the
centroids are deterministic. The recently introduced UCPC approach to
$k$-means clustering~\cite{Gullo:2012:UCPC} is the first work to
acknowledge the importance of probabilistic cluster centroids. However,
it assumes independence in the input and does not support
correlations.

%%%%%%%%%%%%%%%%%%%%%%%%%%%%%%%%%%%%%%%%%%%
{\noindent\bf Data analytics platforms.} Support for iterative
programs is essential in many applications including data mining, web
ranking, graph analysis, and model fitting. This has recently led to a
surge in data-intensive computing platforms with built-in iteration
capability.  Mahout is a library that supports iterative programming
on large clusters~\cite{Mahout:2012}. HaLoop allows iterative
applications to be assembled from existing MapReduce Hadoop
programs~\cite{Haloop2012}. REX supports iterative distributed
computations along data\-base operations in which changes are
propagated between iterations~\cite{Ives:RQL:2012}. MADlib is an
open-source library for scalable in-database analytics~\cite{MADlib:VLDB:2012}. Similar in spirit, Bismarck is a
unified architecture for in-database analytics~\cite{Re:SIGMOD:2012}.
In the Iterative Map-Reduce-Update programming abstraction for machine
learning, user programs are compiled into declarative, optimisable
Datalog code~\cite{DatalogML:Bull:2012}. Platforms that facilitate
uniform treatment of data-intensive tasks were also proposed outside
the data management community, e.g., to support expressive languages
for recursive problems that can be used to automatically synthesise
programs targeting a massively parallel
processor~\cite{Oege:PLDI:2012}.

A key aspect that differentiates ENFrame from the above platforms is
the probabilistic nature of input data and of the whole computation
process. This calls for specifically tailored algorithms. So far,
ENFrame lacks the scalability achievable by the above platforms, since
it only distributes the probability computation task, while the actual
mining task is performed on one machine. The next step is to consider
a fully distributed computational approach.

\nop{Query languages that are analogs of relational algebra or SQL have
been studied for probabilistic
databases~\cite{ISQL:SIGMOD:2007,MCDB:TODS:2011}. Beyond them,
iterating languages with probabilistic changes to the database, such
as probabilistic extensions of fixpoint/while languages, can be used
to declaratively specify queries over Markov Chains, random walks and
stochastic processes~\cite{Koch:MCfixpoint:2010}. A declarative
framework that uses such an expressive language has been proposed for
managing data sourcing games~\cite{Deutch:WWW:2012}, where the
underlying model is that of Markov Chain Monte Carlo.}

%%%%%%%%%%%%%%%%%%%%%%%%%%%%%%%%%%%%%%%%%%%
{\noindent\bf Provenance in database and workflow systems.} To enable
probability computation, we trace fine-grained provenance of the user
computation. This is in line with a wealth of work in probabilistic
databases~\cite{SORK:2011}. Our event language is influenced by work
on provenance semirings~\cite{Green-2007} and
semimodules~\cite{Amsterdamer2011,FHO:VLDB12} that capture provenance
for positive queries with aggregates in relational databases. The
construct $\Phi\otimes v$ resembles the algebraic structure of a
semimodule that is a tensor product of the Boolean semiring
$\mathbb{B}[{\bf X}]$ freely generated by the variable set ${\bf X}$
and of the SUM monoid over the real numbers $\mathbb{R}$. There are
two differences between our construct and these structures. Firstly,
we allow negation in events, which is not captured by the Boolean
semiring. Secondly, even for positive events, $\mathbb{B}[{\bf
  X}]\otimes\mathbb{R}$ is not a semimodule since it violates the
following law: $(\Phi_1 \vee \Phi_2)\otimes v = \Phi_1\otimes v +
\Phi_2\otimes v$. Indeed, under an assignment that maps both $\Phi_1$
and $\Phi_2$ to $\top$, the left side of the equality evaluates to
$v$, whereas the right side becomes $v+v$. Furthermore, our event
language allows to define events via iterations, as needed to
succinctly trace data mining computation.

Workflows employ a much wider variety of programming constructs than
databases. Workflow provenance aims to capture a complete description
of evaluation of a workflow~\cite{Davidson:Bull:2007}, though it sees
tasks as black-boxes, and therefore, consider all outputs of a task to
depend on all of its inputs. This provenance model is too coarse to
support exact derivations of output as needed in our case for
probability computation.

A distinct line of work is on such relational provenance systems as
Perm~\cite{PERM:ICDE:2009}, DBNotes~\cite{DBNotes:VLDBJ:2005},
Orchestra~\cite{Orchestra:SIGREC:2008} that trace provenance using
query rewriting or modified query
operators. Panda~\cite{Panda:BULL:2010} enables provenance-aware
querying.

% Acknowledgements
\smallskip\parhead{Acknowledgements.} This research was supported by EPSRC grant agreement ADEPT (EP/I000194/1).

% This creates a small bibliography, whilst maintaining proper spacing between the heading and the list of citations
\begingroup
 \let\origsection\section
  \renewcommand{\section}[2][]{\origsection[#1]{#2}\ } % yes, that's a deliberate space at the end
  \bibliographystyle{abbrv}
  \begin{small}\bibliography{main}\end{small}
\endgroup

\end{document}